\documentclass[a4paper,twocolumn, 10pt]{IEEEtran} 

\newlength{\figureheight} 
\newlength{\figurewidth} 

\newcommand{\exportFigures}{false}
\newcommand{\arxiv}{true} 

\usepackage{amsthm,amsmath,array}   
\usepackage{amssymb}
\usepackage{graphicx}
\usepackage{color}
\usepackage{xcolor}
\usepackage{cite}
\usepackage{bm}
\usepackage{epsfig,psfrag}
\usepackage{tabularx}
\usepackage{acronym}
\usepackage[yyyymmdd,hhmmss]{datetime}
\usepackage{bbm}
\usepackage{notation}
\usepackage{mathrsfs} 
\usepackage{bardiag}
\usepackage{tabularx}
\usepackage{multirow}
\usepackage{colortbl,pgfplotstable}
\usepackage{dsfont}
\usepackage{times}
\usepackage{rotating}
\usepackage{enumitem}

\usepackage[ruled,vlined,linesnumbered]{algorithm2e}
\SetKwRepeat{Do}{do}{while}

\usepackage[caption=false,font=footnotesize]{subfig}

\usepackage{pgfplots}
\usepackage{tikzscale}
\usepackage{tikz}

\usetikzlibrary{shapes.misc}
\usetikzlibrary{arrows}
\usetikzlibrary{arrows.meta}
\usetikzlibrary{angles}
\usetikzlibrary{quotes}
\usetikzlibrary{fit}
\usetikzlibrary{graphs}
\usetikzlibrary{patterns}
\usetikzlibrary{petri}
\usetikzlibrary{plotmarks}
\usetikzlibrary{positioning}
\usetikzlibrary{decorations.pathmorphing}
\usetikzlibrary{decorations.markings}
\usetikzlibrary{decorations.pathreplacing}
\usetikzlibrary{decorations.shapes}
\usetikzlibrary{decorations.text}
\usetikzlibrary{backgrounds}
\usetikzlibrary{calc}
\usetikzlibrary{spy}
\usetikzlibrary{shapes,arrows}

\usepackage{tikzscale}

\usepackage{pgfplots}
\pgfplotsset{compat=newest}
\usepgfplotslibrary{colormaps}
\usepgfplotslibrary{patchplots}
\usepgfplotslibrary{units}

\usepackage[most,theorems]{tcolorbox}
\tcbuselibrary{most}
\tcbset{every box/.style={enhanced,breakable},shield externalize}

\definecolor{darkgreen}{rgb}{0.0, 0.5, 0.0}
\definecolor{magenta_}{rgb}{1.0, 0, 1.0}

\definecolor{darkgrey}{rgb}{0.3, 0.3, 0.3}

\newcommand{\be}{\begin{equation}}
\newcommand{\ee}{\end{equation}}
\newcommand{\ist}{\hspace*{.3mm}}

\newcommand{\rmv}{\hspace*{-.3mm}}
\newcommand{\iist}{\hspace*{1mm}}
\newcommand{\rrmv}{\hspace*{-1mm}}
\newcommand{\rreq}{\hspace*{-1mm}= \hspace*{-1mm}}
\newcommand{\rrneq}{\hspace*{-1mm} \neq \hspace*{-1mm}}

\newcommand{\nn}{\nonumber}

\makeatletter
\newcommand*\bigcdot{\mathpalette\bigcdot@{.5}}
\newcommand*\bigcdot@[2]{\mathbin{\vcenter{\hbox{\scalebox{#2}{$\m@th#1\bullet$}}}}}
\makeatother

\allowdisplaybreaks
\newcommand{\ticked}{$\text{\rlap{$\checkmark$}}\square$}
\newcommand{\unticked}{{$\square$}}
\newcommand{\tick}[1]{\ifthenelse{#1=1}{\ticked}{\unticked}}

\newcolumntype{L}[1]{>{\raggedright\arraybackslash}p{#1}}
\newcolumntype{C}[1]{>{\centering\arraybackslash}p{#1}}
\newcolumntype{R}[1]{>{\raggedleft\arraybackslash}p{#1}}

\providecommand{\ist}{\hspace*{.3mm}}

\providecommand{\rmv}{\hspace*{-.3mm}}
\providecommand{\iist}{\hspace*{1mm}}
\providecommand{\rrmv}{\hspace*{-1mm}}
\providecommand{\nn}{\nonumber}

\newcommand{\T}{\text{T}}

\definecolor{colA}{rgb}{0,0,1}
\definecolor{colB}{rgb}{0.5,0.0,0.5}
\definecolor{colC}{rgb}{0,0.5,0}
\definecolor{colD}{rgb}{0.1333,0.5451,0.1333}
\definecolor{colE}{rgb}{1.00000,0.55,0.00000}
\definecolor{colF}{rgb}{1.00000,0.0,0.00000}
\definecolor{col_mag}{rgb}{1.00000,0.00000,1.00000}
\definecolor{col_cyan}{rgb}{0,1,1}
\definecolor{col_grey}{rgb}{0.4,0.4,0.4}
\definecolor{col_grey2}{rgb}{0.6,0.6,0.6}
\definecolor{lightgray}{rgb}{0.9,0.9,0.9}  
\definecolor{Gray}{rgb}{225, 225, 225} 
\definecolor{matlabBlue}{rgb}{0.00000,0.44700,0.74100}%
\definecolor{matlabOrange}{rgb}{0.85000,0.32500,0.09800}%
\definecolor{matlabYellow}{rgb}{0.92900,0.69400,0.12500}%
\definecolor{matlabLila}{rgb}{0.49400,0.18400,0.55600}%
\definecolor{matlabGreen}{rgb}{0.46600,0.67400,0.18800}%
\definecolor{col_PAgreen}{rgb}{0.00000,0.49800,0.00000}%

\def\linewidthA{0.5}

\def\mylinewidth{0.7}
\def\mylinewidth2{1}

\def\figureW{6cm}
\def\figureW2{5cm}

\def\marksizeA{1.5}

 \pgfplotsset{stylePA/.style={mymarkfixednumber={\markA}{mark size=\marksizeA,solid}{12},color=matlabBlue, line width=\linewidthA, mark repeat=1, mark phase = 0, mark options={solid,matlabBlue}}}
\pgfplotsset{stylePA_2/.style={mymarkfixednumber={\markA}{mark size=\marksizeA,solid}{6},color=matlabBlue,densely dashed, line width=\linewidthA, mark repeat=1, mark phase = 0, mark options={solid,matlabBlue}}}
\pgfplotsset{stylePA_true/.style={mymarkfixednumber={\markAa}{mark size=\marksizeA,solid}{6},color=matlabBlue,densely dotted, line width=\linewidthA, mark repeat=1, mark phase = 0, mark options={solid,matlabBlue}}}

\pgfplotsset{styleVA1/.style={mymarkfixednumber={\markB}{mark size=\marksizeA,solid}{10},color=matlabOrange, line width=\linewidthA, mark repeat=1, mark phase = 0, mark options={solid,matlabOrange}}}
\pgfplotsset{styleVA1_2/.style={mymarkfixednumber={\markB}{mark size=\marksizeA,solid}{10},color=matlabOrange,densely dashed, line width=\linewidthA, mark repeat=1, mark phase = 0, mark options={solid,matlabOrange}}}
\pgfplotsset{styleVA1_true/.style={mymarkfixednumber={\markBb}{mark size=\marksizeA,solid}{8},color=matlabOrange,densely dotted, line width=\linewidthA, mark repeat=1, mark phase = 0, mark options={solid,matlabOrange}}}

\pgfplotsset{styleVA2/.style={mymarkfixednumber={\markC}{mark size=\marksizeA,solid}{10},color=matlabYellow, line width=\linewidthA, mark repeat=1, mark phase = 0, mark options={solid,matlabYellow}}}
\pgfplotsset{styleVA2_2/.style={mymarkfixednumber={\markC}{mark size=\marksizeA,solid}{10},color=matlabYellow,densely dashed, line width=\linewidthA, mark repeat=1, mark phase = 0, mark options={solid,matlabYellow}}}
\pgfplotsset{styleVA2_true/.style={mymarkfixednumber={\markCc}{mark size=\marksizeA,solid}{8},color=matlabYellow,densely dotted, line width=\linewidthA, mark repeat=1, mark phase = 0, mark options={solid,matlabYellow}}}

\pgfplotsset{styleVA3/.style={mymarkfixednumber={\markD}{mark size=\marksizeA,solid}{10},color=matlabGreen, line width=\linewidthA, mark repeat=1, mark phase = 0, mark options={solid,matlabGreen}}}
\pgfplotsset{styleVA3_2/.style={mymarkfixednumber={\markD}{mark size=\marksizeA,solid}{10},color=matlabGreen,densely dashed, line width=\linewidthA, mark repeat=1, mark phase = 0, mark options={solid,matlabGreen}}}
\pgfplotsset{styleVA3_true/.style={mymarkfixednumber={\markDd}{mark size=\marksizeA,solid}{8},color=matlabGreen,densely dotted, line width=\linewidthA, mark repeat=1, mark phase = 0, mark options={solid,matlabGreen}}}

\pgfplotsset{styleVA4/.style={mymarkfixednumber={\markE}{mark size=\marksizeA,solid}{10},color=matlabLila, line width=\linewidthA, mark repeat=1, mark phase = 0, mark options={solid,matlabLila}}}
\pgfplotsset{styleVA4_2/.style={mymarkfixednumber={\markE}{mark size=\marksizeA,solid}{10},color=matlabLila,densely dashed, line width=\linewidthA, mark repeat=1, mark phase = 0, mark options={solid,matlabLila}}}
\pgfplotsset{styleVA4_true/.style={mymarkfixednumber={\markEe}{mark size=\marksizeA,solid}{8},color=matlabLila,densely dotted, line width=\linewidthA, mark repeat=1, mark phase = 0, mark options={solid,matlabLila}}}

\pgfdeclarelayer{backback1}
\pgfdeclarelayer{back1}
\pgfdeclarelayer{lay1}
\pgfdeclarelayer{backback2}
\pgfdeclarelayer{back2}
\pgfdeclarelayer{lay2}
\pgfsetlayers{backback2,back2,lay2,backback1,back1,lay1,main}

\definecolor{FGgreen}{RGB}{34,139,34}
\definecolor{FGblue}{RGB}{80,120,255}
\definecolor{FGred}{RGB}{255,110,110}

\colorlet{fgColor_node}{colA}
\colorlet{fgColorUp1}{colB}
\colorlet{fgColor_coop}{colB}
\colorlet{fgColorBG}{colC}
\colorlet{fgColor_facAlpha}{colE}
\colorlet{fgColorPost3}{colF}
\colorlet{fgColorComb}{colE}
\colorlet{fgColorBox}{colE}

\newcommand{\mumN}[2]{\ensuremath{\mu_\text{m}\big(\overline{\V{x}}^{(j)}_{#1 , n}\big)}}
\newcommand{\mumL}[2]{\ensuremath{\mu_\text{m}\big(\underline{\V{x}}^{(j)}_{#1 , n}\big)}}

\newcommand{\xn}[1]{\V{x}_{#1}}

\newcommand{\minus}{\rmv - \rmv}

\newcommand{\s}{\hspace*{0.5pt}}

\newcommand{\atan}{\text{atan2}}

\newcommand{\zd}{\ensuremath{{z_\tau^{(j)}}_{\rmv\rmv\rmv\rmv\rmv\rmv\rmv  m,n}}}
\newcommand{\zu}{\ensuremath{{z_\mathrm{u}^{(j)}}_{\rmv\rmv\rmv\rmv\rmv\rmv\rmv m,n}}}

\newcommand{\zaoa}{\ensuremath{{z_\theta^{(j)}}_{\rmv\rmv\rmv\rmv\rmv\rmv\rmv  m,n}}}
\newcommand{\zaod}{\ensuremath{{z_\vartheta^{(j)}}_{\rmv\rmv\rmv\rmv\rmv\rmv\rmv  m,n}}}

\newcommand{\zdr}{\ensuremath{{\rv{z}_\tau^{(j)}}_{\rmv\rmv\rmv\rmv\rmv\rmv\rmv  m,n}}}
\newcommand{\zur}{\ensuremath{{\rv{z}_\mathrm{u}^{(j)}}_{\rmv\rmv\rmv\rmv\rmv\rmv\rmv m,n}}}

\newcommand{\zaoar}{\ensuremath{{\rv{z}_\theta^{(j)}}_{\rmv\rmv\rmv\rmv\rmv\rmv\rmv  m,n}}}
\newcommand{\zaodr}{\ensuremath{{\rv{z}_\vartheta^{(j)}}_{\rmv\rmv\rmv\rmv\rmv\rmv\rmv  m,n}}}

\usetikzlibrary{shapes.misc}
\usetikzlibrary{arrows}
\usetikzlibrary{arrows.meta}
\usetikzlibrary{angles}
\usetikzlibrary{quotes}
\usetikzlibrary{fit}
\usetikzlibrary{graphs}
\usetikzlibrary{patterns}
\usetikzlibrary{petri}
\usetikzlibrary{plotmarks}
\usetikzlibrary{positioning}
\usetikzlibrary{decorations.pathmorphing}
\usetikzlibrary{decorations.markings}
\usetikzlibrary{decorations.pathreplacing}
\usetikzlibrary{decorations.shapes}
\usetikzlibrary{decorations.text}
\usetikzlibrary{backgrounds}
\usetikzlibrary{calc}
\usetikzlibrary{spy}

\usepackage{tikzscale}

\usepackage{pgfplots}
\pgfplotsset{compat=newest}
\usepgfplotslibrary{colormaps}
\usepgfplotslibrary{patchplots}
\usepgfplotslibrary{units}

\def\figHresults{3cm}
\def\figWresults{4.5cm}

\usepackage{ifluatex}

\ifluatex 
    \usepackage{fontspec}
    \defaultfontfeatures{Ligatures={TeX}}

\else
    \usepackage[latin1]{inputenc}
    \usepackage[T2A,T1]{fontenc}
\fi

\ifthenelse{\equal{\exportFigures}{true}}
{
  \usepgfplotslibrary{external}\tikzexternalize[prefix=\tikzfolder]
  \tikzset{external/system call={pdflatex \tikzexternalcheckshellescape -halt-on-error -interaction=batchmode -jobname "\image" "\texsource"}}
  \tikzset{external/up to date check=diff}
}
{}

\def\addlegendimage{\csname pgfplots@addlegendimage\endcsname}
\def\addlegendentry{\csname pgfplots@addlegendentry\endcsname}

\newcommand{\pgfref}[1]
{\ifthenelse{\equal{\exportFigures}{true}}
{\tikzexternalenable\ref{#1}\tikzexternalenable}
{\ref{#1}}}

\pgfplotsset{every axis/.append style={
  label style={font=\normalsize},
  tick label style={font=\scriptsize},
  xticklabel={
   \ifdim \tick pt < 0pt
    \pgfmathparse{abs(\tick)}%
    \llap{$-{}$}\pgfmathprintnumber{\pgfmathresult}
   \else
    \pgfmathprintnumber{\tick}
   \fi}
   }}
 
\usetikzlibrary{decorations.markings}
\makeatletter
\tikzset{
  nomorepostactions/.code={\let\tikz@postactions=\pgfutil@empty},
  mymarkfixednumber/.style n args={3}{decoration={markings,
    mark= between positions 0.1 and 1 step (1/#3)*\pgfdecoratedpathlength with{%
        \tikzset{#2,every mark}\tikz@options
        \pgftransformresetnontranslations
        \pgfuseplotmark{#1}%
      },  
    },
    postaction={decorate},
    /pgfplots/legend image post style={
        mark=#1,mark options={#2},every path/.append style={nomorepostactions}
    },
  },
}
\tikzset{
  nomorepostactions/.code={\let\tikz@postactions=\pgfutil@empty},
  mymarkfixeddistance/.style n args={3}{decoration={markings,
    mark= between positions 0 and 1 step #3cm with{%
        \tikzset{#2,every mark}\tikz@options
        \pgftransformresetnontranslations%
        \pgfuseplotmark{#1}%
      },  
    },
    postaction={decorate},
    /pgfplots/legend image post style={
        mark=#1,mark options={#2},every path/.append style={nomorepostactions}
    },
  },
}
\tikzset{
  nomorepostactions/.code={\let\tikz@postactions=\pgfutil@empty},
  mymark/.style n args={3}{decoration={markings,
    mark= between positions 0 and 1 step 0.75cm with{%
        \tikzset{#2,every mark}\tikz@options
        \pgftransformresetnontranslations%
        \pgfuseplotmark{#1}%
      },  
    },
    postaction={decorate},
    /pgfplots/legend image post style={
        mark=#1,mark options={#2},every path/.append style={nomorepostactions}
    },
  },
}
\makeatother

\def\mylinewidth{0.5}

\def\mymarksize{1.5}
\def\mymarksize2{2.5}
\def\markA{o}
\def\markAa{*}
\def\markB{square}
\def\markBb{square*}
\def\markC{triangle}
\def\markCc{triangle*}
\def\markD{diamond}
\def\markDd{diamond*}
\def\markE{pentagon}
\def\markEe{pentagon*}

\pgfdeclareplotmark{markarrow}
{%
  \pgfpathmoveto{\pgfqpoint{\pgfplotmarksize}{0pt}}
  \pgfpathlineto{\pgfqpointpolar{130}{1.5\pgfplotmarksize}}
  \pgfpathmoveto{\pgfqpoint{\pgfplotmarksize}{0pt}}
  \pgfpathlineto{\pgfqpointpolar{-130}{1.5\pgfplotmarksize}}
  \pgfusepathqstroke
}

\DeclareSymbolFont{sfletters}{OML}{cmbrm}{m}{it}
\DeclareMathSymbol{\svartheta}{\mathord}{sfletters}{"23}

\graphicspath{{./plots/}}
\newcommand{\tikzfolder}{./compiledPlots/}

\tcbset{colback=red!10,colframe=red!10,arc=0.5mm}

\newcommand{\legendref}[1]{\tikzexternaldisable\pgfplotslegendfromname{#1}\tikzexternaldisable}

\acrodefplural{esl}[ESLs]{electronic shelf labels}
\acrodef{aoa}[AOA]{angle-of-arrival}
\acrodef{aod}[AOD]{angle-of-departure}
\acrodef{awgn}[AWGN]{additive white Gaussian noise}
\acrodef{blt}[BLT]{bluetooth}
\acrodef{bp}[BP]{belief propagation}
\acrodef{bw}[BW]{bandwidth}
\acrodef{cdf}[CDF]{cumulative distribution function}
\acrodef{ceda}[CEDA]{channel estimation and detection algorithm}
\acrodef{ci}[CI]{confidence interval}
\acrodef{cl}[CL]{confidence level}
\acrodef{crlb}[CRLB]{Cram\'er-Rao lower bound}
\acrodef{dmc}[DMC]{dense multipath component}
\acrodef{dm}[DM]{dense multipath}
\acrodef{dnr}[DNR]{dense-to-noise ratio}
\acrodef{dps}[DPS]{delay power spectrum}
\acrodef{dut}[DUT]{device under test}
\acrodef{eirp}[EIRP]{equivalent isotropic radiated power}
\acrodef{eot}[EOT]{extended object tracking}
\acrodef{fa}[FA]{false alarm}
\acrodef{glrt}[GLRT]{generalized likelihood ratio test}
\acrodef{iid}[iid]{independent and identically distributed}
\acrodef{lhf}[LHF]{likelihood function}
\acrodef{los}[LOS]{line-of-sight}
\acrodef{lpbo}[LPBO]{legacy \ac{pbo}}
\acrodef{mc}[MC]{Monte Carlo}
\acrodef{mf}[MF]{matched filter}
\acrodef{mie}[MIE]{method of interval estimation}
\acrodef{mimo}[MIMO]{multiple input multiple output}
\acrodef{ml}[ML]{maximum likelihood}
\acrodef{mmse}[MMSE]{minimum mean-square error}
\acrodef{mpc}[MPC]{multipath component}
\acrodef{mp}[MP]{multipath}
\acrodef{MP}[MP]{message passing}
\acrodef{mpslam}[MP-SLAM]{multipath-based simultaneous localization and mapping}
\acrodef{Mpslam}[MP-SLAM]{Multipath-based simultaneous localization and mapping}
\acrodef{mse}[MSE]{mean squared error}
\acrodef{mse}[MSE]{mean squared error}
\acrodef{mva}[MVA]{master \ac{va}}
\acrodef{nlike}[NLIKE]{normalized likelihood}
\acrodef{nlos}[NLOS]{non-\ac{los}}
\acrodef{nnlike}[NNLIKE]{normalized noise-free likelihood}
\acrodef{npbo}[NPBO]{new \ac{pbo}}
\acrodef{olos}[OLOS]{obstructed \ac{los}}
\acrodef{ospa}[OSPA]{optimal subpattern assignment}
\acrodef{mospa}[MOSPA]{mean \ac{ospa}}
\acrodef{pa}[PA]{physical anchor}
\acrodef{pbo}[PBO]{potential bias object}
\acrodef{pcb}[PCB]{printed circuit board}
\acrodef{pcrlb}[PCRLB]{posterior Cram\'er-Rao lower bound}
\acrodef{pdaai}[AIPDA]{amplitude-information \ac{pda}}
\acrodef{pdaf}[PDAF]{probabilistic data association filter}
\acrodef{pda}[PDA]{probabilistic data association}
\acrodef{pdf}[PDF]{probability density function}
\acrodef{pdp}[PDP]{power delay profile}
\acrodef{pmf}[PMF]{probability mass function}
\acrodef{pva}[PVA]{potential \ac{va}}
\acrodef{reb}[REB]{ranging error bound}
\acrodef{rmse}[RMSE]{root mean squared error}
\acrodef{rss}[RSS]{received signal strength}
\acrodef{rv}[RV]{random variable}
\acrodef{sinr}[SINR]{signal-to-interference-plus-noise-ratio}
\acrodef{slam}[SLAM]{simultaneous localization and mapping}
\acrodef{smc}[SMC]{specular multipath component}
\acrodef{snr}[SNR]{signal-to-noise-ratio}
\acrodef{spa}[SPA]{sum-product algorithm}
\acrodef{stdv}[STDV]{standard deviation}
\acrodef{tdoa}[TDOA]{time difference of arrival}
\acrodef{tka}[TKA]{trusted keyless access}
\acrodef{toa}[TOA]{time-of-arrival}
\acrodef{ula}[ULA]{uniform linear array}
\acrodef{ura}[URA]{uniform rectangular array}
\acrodef{ut}[UT]{unscented transform}
\acrodef{uwb}[UWB]{ultra wide band}
\acrodef{va}[VA]{virtual anchor}
\acrodef{zzb}[ZZB]{Ziv-Zakai bound}

\IEEEoverridecommandlockouts

\begin{document}
\frenchspacing

\title{\LARGE 
MIMO Multipath-based SLAM for Non-Ideal Reflective Surfaces}
\author{
Lukas Wielandner\IEEEauthorrefmark{1}\IEEEauthorrefmark{2},
Alexander Venus\IEEEauthorrefmark{1}\IEEEauthorrefmark{2},
Thomas Wilding\IEEEauthorrefmark{1}\IEEEauthorrefmark{2}, 
Klaus Witrisal\IEEEauthorrefmark{1}\IEEEauthorrefmark{2}, 
and
Erik Leitinger\IEEEauthorrefmark{1}\IEEEauthorrefmark{2}
\\[3mm]
\small 
\IEEEauthorrefmark{1}\,Graz University of Technology, \IEEEauthorrefmark{2}\,Christian Doppler Laboratory for Location-Aware Electronic Systems\\[0.5mm]
(e-mail: (lukas.wielandner, a.venus, thomas.wilding, witrisal, erik.leitinger)@tugraz.at)
\vspace*{-6mm}
\thanks{This work was supported in part by the Christian Doppler Research Association; the Austrian Federal Ministry for Digital and Economic Affairs; the National Foundation for Research, Technology, and Development. The project has received funding from the European Union's Horizon 2020 research and innovation programme under grant agreement No 101013425.}
}
\maketitle
\renewcommand{\baselinestretch}{0.96}\small\normalsize 
\begin{abstract}
  \ac{Mpslam} is a well established approach to obtain position information of transmitters and receivers as well as information regarding the propagation environments in future \ac{mimo} communication systems.  Conventional methods for \ac{Mpslam} consider  specular reflections of the radio signals occurring at smooth, flat surfaces, which are modeled by \acp{va} that are mirror images of the \acp{pa}, with each \ac{va} generating a single \ac{mpc}. However, non-ideal reflective surfaces (such as walls covered by shelves or cupboards) cause dispersion effects that violate the \ac{va} model and lead to multiple \acp{mpc} that are associated to a single \ac{va}. In this paper, we introduce a Bayesian particle-based \ac{spa} for \ac{Mpslam} in MIMO communications systems. Our method considers non-ideal reflective surfaces by jointly estimating the parameters of individual dispersion models for each detected surface in delay and angle domain leveraging multiple-measurement-to-feature data association. We demonstrate that the proposed SLAM method can robustly and jointly estimate the positions and dispersion extents of ideal and non-ideal reflective surfaces using numerical simulation.

\end{abstract}  
\renewcommand{\baselinestretch}{1}\small\normalsize 


\acresetall

\vspace*{-2mm}

\section{Introduction}\label{sec:introduction}


\ac{Mpslam} is a promising approach to obtain position information of transmitters and receivers as well as information regarding their propagation environments in future mobile communication systems. 
Usually, specular reflections of radio signals at flat surfaces are modeled by \acp{va} that are mirror images of the \acp{pa} \cite{LeitingerJSAC2015,WitrisalSPM2016,LeitMeyHlaWitTufWin:TWC2019,MenMeyBauWin:JSTSP2019}. The positions of these \acp{va} are unknown. \ac{Mpslam} algorithms can detect and localize \acp{va} and jointly estimate the unkown and time-varying position of mobile agents equipped with transmit or receive antennas \cite{GentnerTWC2016,LeitMeyHlaWitTufWin:TWC2019,MenMeyBauWin:JSTSP2019}. 
The availability of \ac{va} location information makes it possible to leverage multipath propagation of radio signals for the localization of the mobile agent and can thus significantly improve its localization accuracy and robustness.

\vspace*{-1mm}
\subsection{State of the Art}

The proposed algorithm follows the feature-based \ac{slam} approach \cite{DurrantWhyte2006, Dissanayake2001}, i.e., the map is represented by an unknown number of \textit{map features}, whose unknown positions are estimated in a sequential (time-recursive) manner. 
Existing \ac{Mpslam} algorithms consider \acp{va} \cite{LeitMeyHlaWitTufWin:TWC2019, LeiGreWit:ICC2019,MenMeyBauWin:JSTSP2019,KimGraGaoBatKimWym:TWC2020,KimGranSveKimWym:TVT2022} or \acp{mva} \cite{LeiMey:Asilomar2020_DataFusion,LeiVenTeaMey:TSP2023} as features to be mapped.
Most of these methods use estimated parameters related to \acp{mpc} contained in the radio signal, such as distances (which are proportional to delays), \acp{aoa}, or \acp{aod} \cite{MenMeyBauWin:JSTSP2019}. 
These parameters are estimated from the signal in a preprocessing stage \cite{ShutWanJos:CSTA2013, BadHanFle:TSP2017, LiLeiVenTuf:TWC2022, GreLeiWitFle:TWC2024} and are used as ``measurements'' available to the \ac{slam} algorithm. 
Complicating factors in feature-based \ac{slam} are measurement origin uncertainty, i.e., the unknown association of measurements with map features, and the time-varing and unknown number of map features. 
State-of-the-art methods for \ac{Mpslam} are Bayesian estimators that consider these challenges in their joint statistical model leveraging probabilistic data association and the new potential object model  \cite{LeitMeyHlaWitTufWin:TWC2019, LeiGreWit:ICC2019, MenMeyBauWin:JSTSP2019, MeyWilJ21, LiLeiVenTuf:TWC2022}. These methods perform the \ac{spa} on the factor graph representing the resulting high dimensional estimation problem \cite{LeitMeyHlaWitTufWin:TWC2019, LeiGreWit:ICC2019, MenMeyBauWin:JSTSP2019}.

However, conventional \ac{Mpslam} methods are prone to fail in scenarios with non-ideal reflective surfaces \cite{KulmerPIMRC2018, WenKulWitWym:TWC2021} that violate the \ac{va} model. Such scenarios lead to \textit{dispersion effects} (delay dispersion, angular dispersion) in the resulting radio signals. Furthermore, conventional \ac{Mpslam} methods \cite{LeitMeyHlaWitTufWin:TWC2019,LeiGreWit:ICC2019,MenMeyBauWin:JSTSP2019,KimGraGaoBatKimWym:TWC2020,KimGranSveKimWym:TVT2022} assume a feature to generate only a single measurement (point object assumption) \cite{LeitMeyHlaWitTufWin:TWC2019,LeiGreWit:ICC2019}, while the discussed dispersion effects can cause multiple measurements related to a single feature. In \cite{WieVenWilLei:JAIF2023_arxiv}, we proposed a method that models the discussed dispersion effects by combining the \ac{va} model with a uniform dispersion model in delay domain by leveraging multiple-measurement-to-feature data association (developed for extended object tracking \cite{MeyWilJ21}) for \ac{Mpslam}. 
This method can capture imperfections, such as caused by non-calibrated antennas, that result in similar dispersion effects over all map features. However, in realistic scenarios, one often faces {varying} dispersion effects for individual \acp{va} due to rough walls \cite{KulmerPIMRC2018, WenKulWitWym:TWC2021} or walls with cupboards or shelves. 


\vspace*{-1mm}
\subsection{Contributions}

In this paper, we introduce a Bayesian particle-based \ac{spa} for \ac{Mpslam} 
with non-ideal reflective surfaces.
Based on the method from \cite{WieVenWilLei:Fusion2023,WieVenWilLei:JAIF2023_arxiv}, we model non-ideal reflective surfaces by capturing the resulting dispersion effects in measurement domain via a uniform dispersion extent model and probabilistic multiple-measurement-to-feature association \cite{MeyerICASSP2020,MeyWilJ21}. Different to \cite{WieVenWilLei:Fusion2023,WieVenWilLei:JAIF2023_arxiv}, the proposed algorithm can handle varying dispersion for individual map features by jointly inferring dedicated dispersion parameters for each \ac{pa} and \ac{va}.
Furthermore, we extend the work to \ac{mimo} systems by considering amplitude, delay, \ac{aoa}, and \ac{aod} measurements and introducing a novel angular dispersion model, which is captured by jointly inferred angular dispersion parameters. Hence allowing us to treat (nearly) ideal surfaces as well as non-ideal (rough) surfaces in \ac{mimo} setups. 
The key contributions are as follows.
\begin{itemize}
\item We use individual dispersion parameters for each \ac{va} to account for the individual non-ideal reflective surfaces.
\item We extend the work presented in \cite{WieVenWilLei:Fusion2023,WieVenWilLei:JAIF2023_arxiv} to \ac{mimo} systems according to \cite{MenMeyBauWin:JSTSP2019}.
\item We demonstrate based on synthetically generated measurements that the proposed \ac{slam} method robustly associates single and multiple measurements per \ac{va}, hence allowing it to easily treat perfect reflectors (point objects) and rough surfaces (extended objects) in the same manner due to the likelihood function introduced in \cite{WieVenWilLei:JAIF2023_arxiv}. 
\end{itemize}

\subsubsection*{Notation}
Random variables are displayed in sans serif, upright fonts; their realizations in serif, italic fonts. 
Vectors and matrices are denoted by bold lowercase and uppercase letters, respectively. For example, a random variable and its realization are denoted by $\rv x$ and $x$, respectively, and a random vector and its realization 
by $\RV x$ and $\V x$, respectively. 
Furthermore, $\|\V{x}\|$ and ${\V{x}}^{\text T}$ denote the Euclidean norm and the transpose of vector $\V x$, respectively; $\propto$ indicates equality up to a normalization factor;
$f(\V x)$ denotes the \ac{pdf} of random vector $\RV x$; 
$f(\V x | \V y)$ denotes the conditional \ac{pdf} of random vector $\RV x$ conditioned on random vector  $\RV y$.
$\delta(\cdot)$ denotes the Dirac delta function. $\atan(p_y,p_x)$ denotes the four quadrant inverse tangent of a vector $\V{p}=[p_x \iist p_y]^\T$.
The Gaussian \ac{pdf} with respect to $\rv{x}$ is $f_\text{N}(x; \mu , \sigma) = \frac{1}{\sqrt{2\pi} \sigma} e^{\frac{-(x-\mu)^2}{2\,\sigma^2}}$ with mean $\mu$ and standard deviation $\sigma$ \cite{Kay1998}. 
The Gamma \ac{pdf} with respect to $\rv{x}$ is denoted as 
$\mathcal{G}(x;\alpha,\beta) = \frac{1}{\beta^\alpha \Gamma(\alpha)} x^{k-1} e^{-\frac{x}{\beta}}$ where $\alpha$ is the shape parameter, $\beta$ is the scale parameter and $\Gamma(\cdot)$ is the gamma-function. 
Finally, we define the uniform \ac{pdf} $f_\mathrm{U}(x;a,b) = 1/(b-a) {1}_{[a,b]}(x)$.

\section{Geometrical Relations}\label{sec:geometricRel} 

At each time $n$, we consider a mobile agent with state $\V{x}_n = [\V{p}_n^\T\,\ist \V{v}_n^\T]^\T$ at position $\V{p}_n = [p_{x,n} \iist p_{y,n}]^\T$ moving with velocity $\V{v}_n = [v_{x,n} \iist v_{y,n}]^\T$, and $J$ base stations, called \acp{pa}, at known positions $\V{p}_{\mathrm{pa}}^{(j)} = \big[ {p}_{x,\mathrm{pa}}^{(j)} \ist\ist\ist {p}_{y,\mathrm{pa}}^{(j)} \big]^{\mathrm{T}} \rmv\rmv\in \mathbb{R}^2\rmv$, $j \rmv\in\rmv \{1,\ldots,J\}$, where $J$ is assumed to be known, in an environment described by reflective surfaces. 
Both the agent and all \acp{pa} are equipped with antenna arrays. 
The geometry of an antenna array is represented by its array element positions, which are defined for the \acp{pa} arrays by the distances $d_\text{ant}^{(j,h)}$ and the angles $\psi_\text{ant}^{(j,h)}$ relative to the \ac{pa} position $\V{p}_{\mathrm{pa}}^{(j)}$ (with known orientation), and for agent array by distance $d_\text{ant}^{(h)}$ and angle $\psi_\text{ant}^{(h)}$ relative to the agent position $\V{p}_n$ and unknown orientation $\kappa_n$. We assume that the agent array is rigidly coupled with $\V{v}_n$, i.e., the array orientation is given as $\kappa_n = \atan(v_{y,n},v_{x,n})$.

\begin{figure}[t]
	\centering   
	\ifthenelse{\equal{\arxiv}{false}}
	{  
		\tikzsetnextfilename{indoor_environment_MPAs}
		\scalebox{1}{\input{./plots/indoor_environment_MPAs.tex}}
	}
	{ 
		\includegraphics{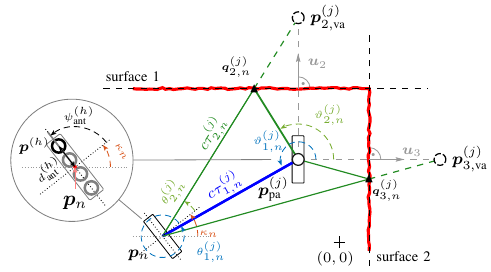} 
	}
	\caption{Exemplary indoor environment including an agent at position $\V{p}_n$, a \ac{pa} at position {$\V{p}^{(j)}_\text{pa} $} and two \acp{va} at positions {$\V{p}^{(j)}_{l,\text{va}}$} for corresponding surfaces. A visualization of the array geometry definition used at the agent and \acp{pa} is included.}
	\vspace*{-4mm}
	\label{fig:overview}
\end{figure}

Specular reflections of radio signals at flat surfaces are modeled by \acp{va} at positions $\V{p}_{l,\mathrm{va}}^{(j)} = \big[ {p}_{x,l,\mathrm{va}}^{(j)} \ist\ist\ist {p}_{y,l,\mathrm{va}}^{(j)} \big]^{\mathrm{T}}$ that are mirror images of \acp{pa} (for details see \cite{LeiVenTeaMey:TSP2023}). 
The according point of reflection $\V{q}_{l,n}^{(j)}$ at the surface is given as
\vspace*{-2mm}
\begin{align}
\V{q}_{l,n}^{(j)} = \V{p}_{l, \text{va}}^{(j)} + \frac{(\V{p}_{\text{pa}}^{(j)} - \V{p}_{l, \text{va}}^{(j)})^\T \V{u}_l}{2(\V{p}_{n} - \V{p}_{l, \text{va}}^{(j)})^\T \V{u}_l} (\V{p}_{n} - \V{p}_{l, \text{va}}^{(j)})
\label{eq:reflectionPoint}\\[-7mm]\nn
\end{align}
where $\V{u}_l$ is the normal vector of the according reflective surface. The point of reflection is needed to relate \ac{aod} measurements of \acp{mpc} to the corresponding \ac{va}. An example is shown in Fig.~\ref{fig:overview}. Since we are using \ac{aod} measurements and \acp{va} as map features, we can only consider \acp{va} associated with a single-bounce path within the algorithm. An extension to multiple-bounce path is possible using the \acp{mva} model \cite{LeiVenTeaMey:TSP2023}.
The current number of \textit{visible} \acp{va} within the scenario is $L_n^{(j)}$ for each of the $J$ \acp{pa}. A \ac{va} does not exist at time $n$, when the path between the agent and this \ac{va} is obstructed. 

\section{Radio Signal Model}\label{sec:signal_model}

At each time $n$, the mobile agent transmits a signal $s(t)$ from an antenna array equipped with $H$ antenna elements and each \ac{pa} $j \rmv\rmv \in \rmv\rmv \{1,\ist\dots\ist,J\}$ acts as a receiver equipped with and array with $H^{(j)}$ antenna elements. In the following, we define $\V{p}^{(j)}_{1,\text{va}} \triangleq \V{p}^{(j)}_\text{pa}$.
%
The received complex baseband signal at the $j$th \ac{pa} is sampled with sampling frequency $ f_{\text{s}} = 1/T_{\text{s}} = B$, giving an observation period of $T =  N_{\text{s}} \, T_{\text{s}}$. 
Stacking these $N_\text{s}$ samples, we obtain the discrete received signal vector in frequency domain between antenna elements $h$ at the agent and $h'$ at \ac{pa} $j$ as \cite{SalVal:JSAC1987,WieVenWilLei:JAIF2023_arxiv}
\vspace*{-2mm}
\begin{align}
\RV{s}_{\text{rx},n}^{(j)}[h,h'] \rmv\rmv&=\rmv\rmv \sum_{l = 1}^{{L}_n^{(j)}}\sum_{i=1}^{{L}_{l,n}^{(j)}} 
\alpha_{l,i,n}^{(j)} 
\mathrm{exp}\big(i2 \pi \frac{f_c}{c} d_\text{ant}^{(h)}\cos(\theta_{l,i,n}^{(j)}\rmv\rmv-\rmv\rmv\psi_\text{ant}^{(h)})\big) \nonumber\\[-1mm] &\quad\times \mathrm{exp}\big(i2 \pi \frac{f_c}{c}d_\text{ant}^{(j,h')}\cos(\vartheta_{l,i,n}^{(j)}\rmv\rmv-\rmv\rmv\psi_\text{ant}^{(j,h')})\big) \nonumber\\ &\quad\times
\V{s}\big(\tau_{l,i,n}^{(j)}\big) + \RV{w}_{n}^{(j)}[h,h']\label{eq:signal_model_sampled}\\[-6mm]\nn
\end{align}
where $[\V{s}(\tau)]_\ell = S(\ell \Delta)\mathrm{exp}(i2\pi \ell\Delta \tau)$ is the $\ell$th frequency sample of the signal spectrum $S(f)$ and $c$ is the speed of light. 
The parameters of the $l$th \ac{mpc} at the agent and the $j$th \ac{pa} (for $l \!=\! 1$) or the corresponding \acp{va} (for $l \rmv\in\rmv \{ 2,\dots,L_n^{(j)}\}$) are the complex amplitude ${\alpha}_{l,i,n}^{(j)} \in \mathbb{C}$, the delay $\tau_{l,i,n}^{(j)}  = \tau(\V{p}_n, \V{p}_{l,\text{va}}^{(j)}) + \nu_{l,i,n}^{(j)}$ with $\tau(\V{p}_n, \V{p}_{l,\text{va}}^{(j)}) \triangleq \big\|\V{p}_n \! -\rmv \V{p}_{l,\text{va}}^{(j)}\big\|/c$, the \ac{aoa} $\theta_{l,i,n}^{(j)} = \theta (\V{x}_{n}, \V{p}_{l,\text{va}}^{(j)}) + \eta_{l,i,n}^{(j)}$ with $\theta (\V{x}_{n}, \V{p}_{l,\text{va}}^{(j)}) \triangleq \atan(p^{(j)}_{{y,l,\text{va}}} \rmv-\rmv p_{y,n}, p^{(j)}_{{x,l,\text{va}}} \rmv-\rmv p_{x,n}  ) \rmv-\rmv \kappa_n$, and \ac{aod} $\vartheta_{l,i,n}^{(j)} = \vartheta(\V{p}_{n}, \V{p}_{l,\text{va}}^{(j)}) + \zeta_{l,i,n}^{(j)}$. For \ac{aod}, we have to distinguish between $\vartheta(\V{p}_{n}, \V{p}_{\text{pa}}^{(j)}) \triangleq \atan(p_{y,n} - p^{(j)}_{{y,\text{pa}}}, p_{x,n} \rmv-\rmv p^{(j)}_{{x,\text{pa}}})$ for  $l=1$  and $\vartheta(\V{p}_{n}, \V{p}_{l,\text{va}}^{(j)}) \triangleq \atan(q^{(j)}_{{y,l,n}} \rmv-\rmv p^{(j)}_{{y,l,\text{va}}}, q^{(j)}_{{x,l,n}} \rmv-\rmv p^{(j)}_{{x,1,\text{va}}})$ for $l\neq 1$, where $\V{q}_{l,n}$ is calculated based on \eqref{eq:reflectionPoint} using $\V{p}_{n}$.

Due to the effects of non-ideal reflective surfaces, there are $L_{l,n}^{(j)}$ sub-components associated to the $l$th \ac{mpc} generated by a marked Poisson point process \cite{SalVal:JSAC1987, PedersenJTAP2018, WieVenWilLei:JAIF2023_arxiv}. Note that with the intensity distribution of the complex amplitudes ${\alpha}_{l,i,n}^{(j)}$ and arrival distributions of the sub-components in delay $\nu^{(j)}_{l,i,n}$, in \ac{aoa} $\eta_{l,i,n}^{(j)}$, and in \ac{aod} $\zeta_{l,i,n}^{(j)}$ many different non-ideal effects of surfaces such as roughness \cite{KulmerPIMRC2018,WenKulWitWym:TWC2021} can be described. However, for the sake of simplicity and effectiveness, we have chosen the following simple model. We assume that for each \ac{mpc} $l$ there is a main-component $i=1$ and the ${L}_{l,n}^{(j)}-1$ sub-components with index $i \in \{2,\dots,{L}_{l,n}^{(j)}\}$. The main-component is described by the complex amplitude ${\alpha}_{l,1,n}^{(j)} \in \mathbb{C}$, the delay $\tau_{l,1,n}^{(j)}  = \tau(\V{p}_n, \V{p}_{l,\text{va}}^{(j)})$, the \ac{aoa} $\theta_{l,1,n}^{(j)} = \theta (\V{x}_{n}, \V{p}_{l,\text{va}}^{(j)})$, and \ac{aod} $\vartheta_{l,1,n}^{(j)} = \vartheta(\V{p}_{n}, \V{p}_{l,\text{va}}^{(j)})$. The sub-components are assumed to have complex amplitudes ${\alpha}_{l,i,n}^{(j)} = {\alpha}_{l,1,n}^{(j)}\beta_{l,i,n}$ with attenuation $\beta_{l,i,n}$ and to be uniformly distributed over the dispersion domains $\nu^{(j)}_{l,i,n} \in [0, \psi^{(j)}_{\tau,l,n}]$, $\eta^{(j)}_{l,i,n} \in [-\psi^{(j)}_{\theta,l,n}/2, \psi^{(j)}_{\theta,l,n}/2]$, and $\zeta^{(j)}_{l,i,n} \in [-\psi^{(j)}_{\vartheta,l,n}/2, \psi^{(j)}_{\vartheta,l,n}/2]$, where $\psi^{(j)}_{\tau,l,n}$, $\psi_{\theta,l,n}^{(j)}$, and $\psi_{\vartheta,l,n}^{(j)}$ are called dispersion parameters. Note that the distributions of these parameters will depend on the environment properties and can vary per component and \ac{pa}.

The noise vector $\RV{w}_{n}^{(j)}[h,h'] \in \mathbb{C}^{N_\text{s}} $ is a zero-mean, circularly-symmetric complex Gaussian random vector with covariance matrix ${\sigma}^{(j)\ist 2} \M{I}_{N_\text{s}}\,\forall\,h,h'$ and noise variance ${\sigma}^{(j)\ist 2}$.

\subsection{Parametric Channel Estimation}  \label{sec:channel_estimation}

By applying at each time $ n $, a \ac{ceda} \cite{ShutWanJos:CSTA2013, BadHanFle:TSP2017, LiLeiVenTuf:TWC2022, GreLeiWitFle:TWC2024} to the full {observed} discrete signal vector $\V{s}_{\text{rx},n}^{(j)}=\big[\V{s}_{\text{rx},n}^{(j)}[1,1]^\T \cdots \V{s}_{\text{rx},n}^{(j)}[H,H^{(j)}]^\T\big]^\T \in\mathbb{C}^{NHH^{(j)}]}$ of all antenna elements at agent and \ac{pa} $j$, one obtains, for each \ac{pa} $j$, a number of $M_n^{(j)}$ measurements denoted by ${\V{z}^{(j)}_{m,n}}$ with $m \in  \Set{M}_n^{(j)} \triangleq \{1,\,\dots\,,M_n^{(j)}\} $.
Each $\V{z}^{(j)}_{m,n} = [\zd \ \zaoa \ \zaod \ \zu]^\text{T}$ representing a potential \ac{mpc} parameter estimate, contains a delay measurement $\zd \rmv\rmv\in\rmv\rmv [0, \tau_\text{max}]$ an \ac{aoa} measurement $\zaoa \rmv\rmv\in\rmv\rmv [-\pi, \pi]$ and an \ac{aod} measurement $\zaod \rmv\rmv\in\rmv\rmv [-\pi, \pi]$. We also get a normalized amplitude measurement\footnote{The normalized amplitude measurements are determined as $\zu  = |\mu^{(j)}_{\alpha,m,n}|/\sigma^{(j)}_{\alpha,m,n}$ with $\mu^{(j)}_{\alpha,m,n} \,\,\, {\in \mathbb{C}}$ and $\sigma^{(j)}_{\alpha,m,n} \,\,\, {\in \mathbb{R}^{+}}$ denoting respectively the estimated mean and standard deviation of the complex amplitudes provided by the \ac{ceda}.} $\zu \rmv\rmv\in\rmv\rmv [\gamma, \infty )$, where $\gamma$ is the detection threshold.
The \ac{ceda} decomposes the signal $\V{s}_{\text{rx},n}^{(j)}$ into individual, decorrelated components reducing significantly the number of dimensions (${M}_n^{(j)} \ll N_\text{s}HH'$). It thus compresses the information contained in $\V{s}_{\text{rx},n}^{(j)}$ into $\vspace*{-1mm}\V{z}^{(j)}_{n} = [{\bm{z}^{(j)\text{T}}_{1,n}}  \rmv \cdots  {\V{z}^{(j)\text{T}}_{M_n^{(j)},n}}]^\text{T}$. The stacked vector $\V{z}_n = [\V{z}^{(1)\, \text{T}}_{n} \rmv \cdots  \V{z}^{(J)\,\text{T}}_{n}]^\text{T}$ is used by the proposed algorithm as a noisy measurement.


\section{System Model}\label{sec:system_model}

At each time $n$, the state $\RV{x}_n = [\RV{p}_n^\T\,\ist \RV{v}_n^\T]^\T$ of the agent consists of its position $\RV{p}_n$ and velocity $\RV{v}_n$.  In line with \cite{MeyerProc2018, LiLeiVenTuf:TWC2022, LeiGreWit:ICC2019}, we account for the unknown number of \acp{va} by introducing for each \ac{pa} $j$ \acp{pva} $k \rmv\in\rmv \mathcal{K}^{(j)}_n \triangleq \{ 1,\dots,{K}_n^{(j)} \}$. 
The number of \acp{pva} $K_n^{(j)}$ is the maximum possible number of \acp{va} of \ac{pa} $j$ that produced measurements so far \cite{MeyerProc2018} (i.e., $K_n^{(j)}$ increases with time). 
The state of \ac{pva} $(j,k)$ is denoted as $\RV{y}_{k,n}^{(j)} \!\triangleq\rmv \big[\RV{x}_{k,n}^{(j)\text{T}}\,\ist \rv{r}_{k,n}^{(j)}\big]^\T$ with $\RV{x}_{k,n}^{(j)} = \big[\RV{p}_{k,\text{va}}^{(j)\text{T}} \,\RV{\psi}^{(j)\T}_{k,n} \big]^\T$, which includes the dispersion parameters $\RV{\psi}^{(j)}_{k,n} = [\rv{\psi}^{(j)}_{\tau,k,n} \iist \rv{\psi}^{(j)}_{\theta,k,n} \iist \rv{\psi}^{(j)}_{\vartheta,k,n}]^\T$. 
The existence/nonexistence of \ac{pva} $k$ is modeled by the existence variable $\rv{r}^{(j)}_{k,n} \rmv\in \{0,1\}$ in the sense that \ac{pva} $k$ exists if and only if $r^{(j)}_{k,n} \!=\! 1$. 
The \ac{pva} state is considered formally also if \ac{pva} $k$ is nonexistent, i.e., if $r^{(j)}_{k,n} \!=\rmv 0$. 
Since a part of the \ac{pa} state is unknown, we also consider the \ac{pa} itself a \ac{pva}. 
Hence, we distinguish between the \ac{pva} $k=1$ that explicitly represents the \ac{pa}, which is a-priori existent and has known (fixed) position $\V{p}_{1,\text{va}}^{(j)} = \V{p}_{\text{pa}}^{(j)} $, and all other \acp{pva} $k \in \{2,\ist\dots\ist, K_n^{(j)}\}$ with unknown existences and positions. 
Note that the \acp{pva} state representing the \ac{pa} still considers the existence variable $\rv{r}^{(j)}_{1,n}$.
The states $\RV{x}^{(j)\ist\text{T}}_{k,n}$ of nonexistent \acp{pva} are obviously irrelevant. 
Therefore, all \acp{pdf} defined for \ac{pva} states, $f(\V{y}_{k,n}) =\rmv f(\V{x}_{k,n}, r_{k,n})$, are of the form $f(\V{x}^{(j)}_{k,n}, 0 )$ $=\rmv f^{(j)}_{k,n} f_{\text{d}}(\V{x}^{(j)}_{k,n})$, where $f_{\text{d}}(\V{x}^{(j)}_{k,n})$ is an arbitrary ``dummy'' \ac{pdf} and $f^{(j)}_{k,n} \!\rmv\in [0,1]$ is a constant. 
We also define the stacked vectors $\RV{y}_n^{(j)} \!\triangleq \big[\RV{y}_{1,n}^{(j)\text{T}} \rmv\cdots\ist \RV{y}_{K_n^{(j)}\rmv,n}^{(j)\text{T}} \big]^\T$ and $\RV{y}_n \!\triangleq \big[\RV{y}_n^{(1)\text{T}} \rmv\cdots\ist \RV{y}_n^{(J)\text{T}} \big]^\T\rmv$. 

\subsection{State Evolution}
\label{sec:state_statistics}

For each \ac{pva} with state $\RV{y}_{k,n-1}^{(j)}$ with $k \rmv\in\rmv \mathcal{K}^{(j)}_{n-1} \triangleq \{ 1,\dots,{K}_{n-1}^{(j)} \}$ at time $n-1$ and \ac{pa} $j$, there is one ``legacy'' \ac{pva} with state $\underline{\RV{y}}_{k,n}^{(j)} \!\triangleq \big[\underline{\RV{x}}_{k,n}^{(j)\text{T}}\,\ist \underline{\rv{r}}_{k,n}^{(j)}\big]^\T$ with $k \in \mathcal{K}^{(j)}_{n-1}$ at time $n$ and \ac{pa} $j$. 
We also define the joint states \vspace*{-0.8mm} $\underline{\RV{y}}_n^{(j)} \triangleq \big[\underline{\RV{y}}_{1,n}^{(j)\text{T}} \rmv\cdots\ist \underline{\RV{y}}_{K_{n-1}^{(j)}\rmv,n}^{(j)\text{T}} \big]^\T\rmv$ and $\underline{\RV{y}}_n \!\triangleq \big[\underline{\RV{y}}_n^{(1)\text{T}} \rmv\cdots\ist \underline{\RV{y}}_n^{(J)\text{T}} \big]^\T\rmv$. 
Assuming that the agent state as well as the \ac{pva} states of all \acp{pa} evolve independently across $k$, $n$, and $j$, the joint state-transition \ac{pdf} factorizes as \cite{LeitMeyHlaWitTufWin:TWC2019,MeyerProc2018}
\vspace*{-2mm}
\begin{align}
	&f\big(\V{x}_n,\underline{\V{y}}_n|\V{x}_{n-1},\V{y}_{n-1}\big) \nn\\[-2mm]
	&\hspace*{12mm}= \rmv\rmv f(\V{x}_{n}|\V{x}_{n-1}) \prod_{j=1}^J \rmv\rmv \rmv\rmv \prod_{k=1}^{K_{n-1}^{(j)}} \rmv\rmv \rmv\rmv f\big(\underline{\V{y}}_{k,n}^{(j)} \big| \V{y}_{k, n-1}^{(j)}\big) 
	\label{eq:state_space}\\[-7mm]\nn
\end{align}
where $f(\underline{\V{y}}_{k,n}^{(j)}| \V{y}_{k, n-1}^{(j)}) \triangleq f\big(\underline{\V{x}}_{k,n}^{(j)},\underline{r}_{k,n}^{(j)} \big| \V{x}_{k,n-1}^{(j)}, r_{k,n-1}^{(j)}\big)\vspace*{0.5mm}$ is the legacy \ac{pva} state-transition \ac{pdf}. 
If \ac{pva} did not exist at time $n \rmv-\! 1$, i.e., $r_{k,n-1}^{(j)} \!=\! 0$, it cannot exist as a legacy \ac{pva} at time $n$ either. 
Thus,
\vspace*{-2mm}
\begin{align}
	f\big(&\underline{\V{x}}_{k,n}^{(j)},r_{k,n}^{(j)} \big| \V{x}_{k,n-1}^{(j)}, 0\big) = \begin{cases} 
		f_\text{d}\big(\underline{\V{x}}_{k,n}^{(j)}\big) , &\!\!\! \underline{r}_{k,n}^{(j)} \!=\rmv 0 \\[0mm]
		0 , &\!\!\! \underline{r}_{k,n}^{(j)} \!=\! 1.
	\end{cases} \!\!\!\!
	\label{eq:dummy_transition}\\[-7mm]\nn
\end{align}
If \ac{pva} existed at time $n \rmv-\! 1$, i.e., $r_{k,n-1}^{(j)} \!=\! 1$, it either dies, i.e., $\underline{r}_{k,n}^{(j)} \!=\rmv 0$, or survives, i.e., $\underline{r}_{k,n}^{(j)} \!=\! 1$ with survival probability denoted as $p_\text{s}$. 
If it does survive, its state $\underline{\RV{y}}_{k,n}^{(j)}$ is distributed according to the state-transition PDF \vspace*{0.2mm}$f\big(\underline{\V{x}}_{k,n}^{(j)}\big| \V{x}_{k,n-1}^{(j)}\big) \triangleq  \delta\big(\underline{\V{p}}_{k,\text{va}}^{(j)} - \V{p}_{k,\text{va}}^{(j)}\big) f\big( \V{\psi}_{k,n}^{(j)} \big| \V{\psi}_{k,n-1}^{(j)}\big)$ \cite{LeiGreWit:ICC2019,LeitMeyHlaWitTufWin:TWC2019}. 
Thus,
\vspace*{-2mm}
\begin{align}
	&f\big(\underline{\V{x}}_{k,n}^{(j)},\underline{r}_{k,n}^{(j)} \big| \V{x}_{k,n-1}^{(j)}, 1\big)\nn\\
	& \hspace{4mm} =\begin{cases} 
		(1 \!-\rmv p_\text{s}) \ist f_\text{d}\big(\underline{\V{x}}_{k,n}^{(j)}\big) , &\!\!\! \underline{r}_{k,n}^{(j)} \!=\rmv 0 \\[0mm]
		p_\text{s} \ist \delta\big(\underline{\V{p}}_{k,\text{va}}^{(j)} - \V{p}_{k,\text{va}}^{(j)}\big) f\big( \V{\psi}_{k,n}^{(j)} \big| \V{\psi}_{k,n-1}^{(j)}\big) , &\!\!\! \underline{r}_{k,n}^{(j)} \!=\! 1
	\end{cases}\ist.
	\label{eq:survival_transition}\\[-6mm]\nn
\end{align}
The agent state $\RV{x}_n$ with state-transition \ac{pdf} $f(\V{x}_{n}|\V{x}_{n-1})$ is assumed to evolve in time according to a 2-dimensional, constant velocity and stochastic acceleration model \cite{BarShalom2002EstimationTracking} given as $\RV{x}_n = \bm{A}\, \RV{x}_{n\minus 1} + \bm{B}\, \RV{w}_{n}$, with the acceleration process $\RV{w}_n$ being \ac{iid} across $n$, zero mean, and Gaussian with covariance matrix ${\sigma_{\text{w}}^2}\, \bm{I}_2$, ${\sigma_{\text{w}}}$ is the acceleration standard deviation, and $\bm{A} \in \mathbb{R}^{4\times 4}$ and $\bm{B} \in \mathbb{R}^{4\times 2}$ are defined according to \cite[p.~273]{BarShalom2002EstimationTracking}, with observation period $\Delta T$. The state-transition \acp{pdf} of the dispersion parameter states $f(\V{\psi}^{(j)}_{k,n}|\V{\psi}^{(j)}_{k,n-1}) = f({\psi}^{(j)}_{\tau,k,n}|\psi^{(j)}_{\tau,k,n-1})f(\psi^{(j)}_{\theta,k,n}|\psi^{(j)}_{\theta,k,n-1})f(\psi^{(j)}_{\vartheta,k,n}|\psi^{(j)}_{\vartheta,k,n-1})$ are assumed to evolve independently of each other across $n$. 
We model the individual state-transition \acp{pdf} by Gamma \acp{pdf} given respectively by $f(\psi^{(j)}_{\tau,k,n}|\psi^{(j)}_{\tau,k,n-1})= \mathcal{G}(\psi^{(j)}_{\tau,k,n};q_{\tau},\psi^{(j)}_{\tau,k,n-1}/q_{\tau})$, $f(\psi^{(j)}_{\theta,k,n}|\psi^{(j)}_{\theta,k,n-1}) = \mathcal{G}(\psi^{(j)}_{\theta,k,n};q_{\theta},\psi^{(j)}_{\theta,k,n-1}/q_{\theta})$ and $f(\psi^{(j)}_{\vartheta,k,n}|\psi^{(j)}_{\vartheta,k,n-1}) = \mathcal{G}(\psi^{(j)}_{\vartheta,k,n};q_{\vartheta},\psi^{(j)}_{\vartheta,k,n-1}/q_{\vartheta})$, where $q_{\tau}$, $q_{\theta}$ and $q_{\vartheta}$ represent the respective state noise parameters \cite{Koc:TAES2008_EOT,MeyWilJ21}. 
Small $q$ implies a large state transition uncertainty.

\subsection{Measurement Model}
\label{sec:measurementModel} 

Before the measurements are observed, they are random and represented by the vector ${\RV{z}^{(j)}_{m,n}} = [\zdr\, \zaoar \, \zaodr \,\zur]^\text{T}$. In line with Section~\ref{sec:channel_estimation} we define the nested random vectors $\RV{z}^{(j)}_{n} = [{\RV{z}^{(j)\text{T}}_{1,n}} \cdots\ist {\RV{z}^{(j)\text{T}}_{\rv{M}_n^{(j)},n}}]^\text{T}$, with length corresponding to the random number of measurements $\rv{M}^{(j)}_n$, and $\RV{z}_n = [\RV{z}^{(1)\, \text{T}}_{n}\cdots\ist \RV{z}^{(J)\,\text{T}}_{n}]^\text{T}$.
The vector containing all numbers of measurements is defined as  $\RV{M}_n = [\rv{M}_n^{(1)}\ist\cdots\ist \rv{M}_n^{(J)}]^\text{T}$.

If \ac{pva} $k$ exists ($r^{(j)}_{k,n} = 1$), it gives rise to a random number of measurements. 
The mean number of measurements per (existing) \ac{pva} is modeled by a Poisson point process with mean $\mu_\text{m}\big(\V{\psi}^{(j)}_{k,n} \big)$. The individual measurements $\RV{z}^{(j)}_{m,n} $ are assumed to be conditionally independent, i.e., the joint \ac{pdf} of all measurements factorizes as $f(\V{z}_{n}^{(j)}| {M}_n^{(j)}, {\V{x}}_n, {\nu}_{k,n}^{(j)}, {\eta}_{k,n}^{(j)}, {\zeta}_{k,n}^{(j)}, \V{x}_{k,n}^{(j)})  = \prod_{m=1}^{{M}_n^{(j)}} f(\V{z}_{m,n}^{(j)}| {\V{x}}_n, {\nu}_{k,n}^{(j)}, {\eta}_{k,n}^{(j)}, {\zeta}_{k,n}^{(j)}, \V{x}_{k,n}^{(j)}) $.

If ${\RV{z}^{(j)}_{m,n}}$ is generated by a \ac{pva}, i.e., it corresponds to a main-component (\ac{los} component or \ac{mpc}),
we assume that the single-measurement \ac{lhf} $f(\V{z}_{m,n}^{(j)}| {\V{x}}_n,  {\nu}_{k,n}^{(j)}, {\eta}_{k,n}^{(j)}, {\zeta}_{k,n}^{(j)}, \V{x}_{k,n}^{(j)})$ is conditionally independent across $\zdr$, $\zaoar$ and $\zaodr$. 
Thus, it factorizes as
\vspace{-2mm}
\begin{align}
f(\V{z}_{m,n}^{(j)}| {\V{x}}_n,  \V{x}_{k,n}^{(j)}, \nu_{k,n}^{(j)}, \eta_{k,n}^{(j)},\zeta_{k,n}^{(j)}) &  = f(\zd| \V{p}_{n}, \V{x}_{k,n}^{(j)}, {\nu}_{k,n}^{(j)}) \nonumber\\
	& \hspace{-5cm} \times f(\zaoa|\V{x}_{n}, \V{x}_{k,n}^{(j)},{\eta}_{k,n}^{(j)}) f(\zaod|\V{p}_{n}, \V{x}_{k,n}^{(j)},{\zeta}_{k,n}^{(j)}).
	\label{eq:single_measurement_likelihood}\\[-6mm]\nn
\end{align}
The \ac{lhf} of the corresponding delay measurement $\zdr$ is given by
\vspace*{-1mm}
\begin{align}
	&  f(\zd | \V{p}_{n}, \V{x}_{k,n}^{(j)}, {\nu}_{k,n}^{(j)}) \nonumber\\
	&\hspace{10mm} = f_\text{N}\Big(\zd;\, \tau( \V{p}_{n}, \V{p}_{k,\text{va}}^{(j)}) + \nu_{k,n}^{(j)},\, \sigma_{\tau}^{2} (\zu) \Big)\\[-7mm]\nn
\end{align}
with mean $\tau( \RV{p}_{n}, \RV{p}_{k,\text{va}}^{(j)}) + \rv{\nu}_{k,n}^{(j)}$ and variance $\sigma_{\tau}^{2} ( \zu )$. 
The standard deviation is determined from the Fisher information given by $\sigma_{\tau}^{2} (u) = c^2 / ( 8\, \pi^2 \, \beta_\text{bw}^2 \, u^{2})$ with $\beta_\text{bw}$ being the root mean squared bandwidth \cite{WilGreLeiMueWit:ACSSC2018} (see Section~\ref{sec:results}). 
The \ac{lhf} of the corresponding \ac{aoa} measurement $\zaoa$ is obtained as
\vspace*{-1mm}
\begin{align} \label{eq:lhf_aoa}
	&  f(\zaoa | \V{x}_{n}, \V{x}_{k,n}^{(j)}, {\eta}_{k,n}^{(j)}) \nonumber\\
	&\hspace{10mm} = f_\text{N}\Big(\zaoa;\, \theta( \V{x}_{n}, \RV{p}_{k,\text{va}}^{(j)}) + \eta_{k,n}^{(j)},\, \sigma_{\theta}^{2} (\zu) \Big)\\[-7mm]\nn
\end{align}
with mean $\theta( \RV{x}_{n}, \RV{p}_{k,\text{va}}^{(j)}) + \rv{\eta}_{k,n}^{(j)}$ and variance $\sigma_{\theta}^{2} ( \zu )$ \cite{WilGreLeiMueWit:ACSSC2018}.
The \ac{lhf} of the corresponding \ac{aod} measurement $\zaod$ is defined in a similar manner.
Based on the dispersion model introduced in the Section~\ref{sec:signal_model}, the joint \ac{pdf} of the dispersion variables can be constructed as follows \cite{WieVenWilLei:JAIF2023_arxiv}
\vspace*{-1mm}
\begin{align} \label{eq:dispersion}
	f( \nu_{k,n}^{(j)}, {\eta}_{k,n}^{(j)}, {\zeta}_{k,n}^{(j)}|  \V{\psi}_{n} ) =& \frac{1}{2} \left(\delta(\nu_{k,n}^{(j)})\, \delta({\eta}_{k,n}^{(j)}), \delta({\zeta}_{k,n}^{(j)})\right. \nonumber \\
	& \hspace{-3.5cm} +   \delta(\nu_{k,n}^{(j)} - {\psi}^{(j)}_{\tau,k,n})\, \delta({\eta}_{k,n}^{(j)}-{\psi}^{(j)}_{\theta,k,n}),  \delta({\zeta}_{k,n}^{(j)}-{\psi}^{(j)}_{\vartheta,k,n})  \nn \\
	&  \hspace{-3.5cm} \times  f_\text{U} (\nu_{k,n}^{(j)};0, {\psi}^{(j)}_{\tau,k,n}) f_\text{U}(\eta_{k,n}^{(j)};-\psi^{(j)}_{\theta,k,n}/2, \psi^{(j)}_{\theta,k,n}/2)
	\nn \\
	&  \hspace{-3.5cm} \times \left.  f_\text{U}(\zeta_{k,n}^{(j)};-\psi^{(j)}_{\vartheta,k,n}/2, \psi^{(j)}_{\vartheta,k,n}/2) \right)\\[-7mm]\nn
\end{align}
where the according delay dispersion random variable is given as $\rv{\nu}_{k,n}^{(j)} \sim  f_\text{U}({\nu}_{k,n}^{(j)};0,{\psi}^{(j)}_{\tau,k,n})$ and the angular dispersion random variables are $\rv{\eta}_{k,n}^{(j)} \sim f_\text{U}({\eta}_{k,n}^{(j)};-{\psi}^{(j)}_{\theta,k,n}/2,{\psi}^{(j)}_{\theta,k,n}/2)$ and $\rv{\zeta}_{k,n}^{(j)} \sim f_\text{U}({\zeta}_{k,n}^{(j)};-{\psi}^{(j)}_{\vartheta,k,n}/2,{\psi}^{(j)}_{\vartheta,k,n}/2)$. The \ac{pdf} of a single measurement $\RV{z}_{m,n}^{(j)}$ can now be obtained by integrating out the dispersion variables as
\vspace*{-1mm}
\begin{align}
f(\V{z}_{m,n}^{(j)}| \V{x}_n,  \V{x}_{k,n}^{(j)}) =&  \int 	f(\V{z}_{m,n}^{(j)}| \V{x}_n,  \V{p}_{k,\text{va}}^{(j)}, {\nu}_{k,n}^{(j)}, \eta_{k,n}^{(j)}, \zeta_{k,n}^{(j)}) \nonumber\\
	& \hspace{-2cm} \times f( \nu_{k,n}^{(j)}, \eta_{k,n}^{(j)}, \zeta_{k,n}^{(j)} |  \V{\psi}^{(j)} _{k,n} ) \mathrm{d}\nu_{k,n}^{(j)}\, \mathrm{d}{\eta}_{k,n}^{(j)} \mathrm{d}{\zeta}_{k,n}^{(j)}  \nonumber \\
	 & \hspace{-2.5cm} = f(\zd | \V{p}_{n}, \V{p}_{k,\text{va}}^{(j)})   	f(\zaoa| \V{x}_{n}, \V{p}_{k,\text{va}}^{(j)} ) f(\zaod| \V{p}_{n}, \V{p}_{k,\text{va}}^{(j)} )   \nonumber \\
	&	\hspace{-2cm}  + 		f(\zd| \V{p}_{n},  \V{p}_{k,\text{va}}^{(j)}, \psi_{\tau, k}^{(j)})  f(\zaoa| \V{x}_{n}, \V{p}_{k,\text{va}}^{(j)}, \psi_{\theta, k}^{(j)})  \nonumber \\ 
	& \hspace{-2cm}  \times f(\zaod| \V{p}_{n}, \V{p}_{k,\text{va}}^{(j)}, \psi_{\vartheta, k}^{(j)})  
\end{align}
with the main-component \acp{pdf}
\begin{align}
\hspace{-1cm} f(\zd | \V{p}_{n}, \V{p}_{k,\text{va}}^{(j)})  \rrmv & = \rrmv f_\text{N}(\zd; \rmv \tau( \V{p}_{n}, \V{p}_{k,\text{va}}^{(j)})  \rmv ,  \rmv \sigma_{\tau}^{2} (\zu)\\
f(\zaoa | \V{x}_{n}, \V{p}_{k,\text{va}}^{(j)})  \rrmv & =  \rrmv f_\text{N}(\zaoa; \rmv  \theta (\V{x}_{n}, \V{p}_{k,\text{va}}^{(j)}) \rmv ,  \rmv \sigma_{\theta}^{2} (\zu)\\
f(\zaod | \V{p}_{n}, \V{p}_{k,\text{va}}^{(j)})  \rrmv & =  \rrmv f_\text{N}(\zaod; \rmv  \vartheta( \V{p}_{n}, \V{p}_{k,\text{va}}^{(j)}) \rmv , \rmv \sigma_{\vartheta}^{2} (\zu) \label{eq:mainAoD}
\end{align}
as well as the additional sub-component \acp{pdf}
\vspace{-2mm}
\begin{align}
	& 	f(\zd | \V{p}_{n}, \V{p}_{k,\text{va}}^{(j)},{\psi}^{(j)}_{\tau,n}) \nonumber \\
	&= \rmv\rmv\frac{1}{{\psi}^{(j)}_{\tau,n}} \rmv\rmv \int_{0}^{{\psi}^{(j)}_{\tau,n}} \rmv  \rmv \rmv \rmv \rmv \rmv \rmv\rmv\rmv\rmv\rmv\rmv\rmv\rmv\rmv\rmv\rmv
	f_\text{N}\Big(\zd; \tau(\V{p}_n, \V{p}_{k,\text{va}}^{(j)}) \rmv\rmv + \rmv\rmv \nu_{k,n}^{(j)}, \sigma_{\tau}^{2} (\zu) \rmv\rmv \Big)   \s \mathrm{d} \nu_{\rmv  k,n}^{(j)} \nonumber \\	
	&= \frac{1}{2 {\psi}^{(j)}_{\tau,n}} \Bigg( \text{erf}\left( \frac{\tau(\V{p}_n, \V{p}_{k,\text{va}}^{(j)}) + {\psi}^{(j)}_{\tau,n} - \zd }{  \sigma_{\tau} (\zu) \sqrt{2}} \right) 
	\nonumber \\ &\hspace{25mm} - \text{erf}\Bigg( \frac{\tau(\V{p}_n, \V{p}_{k,\text{va}}^{(j)}) - \zd }{  \sigma_{\tau} ( \zu) \sqrt{2}} \Bigg) \Bigg)\\[-7mm]\nn
\end{align}
and
\vspace{-1mm}
\begin{align}
& 	f(\zaoa | \V{x}_{n}, \V{p}_{k,\text{va}}^{(j)},\psi^{(j)}_{\theta,n}) \nonumber \\
	&= \rmv\rmv\frac{1}{{\psi}^{(j)}_{\theta,n}} \rmv\rmv\rmv\rmv \int_{-{\psi}^{(j)}_{\theta,n}/2}^{\psi^{(j)}_{\theta,n}/2} \rmv  \rmv \rmv \rmv \rmv \rmv \rmv\rmv\rmv\rmv\rmv\rmv\rmv\rmv\rmv\rmv\rmv
	f_\text{N}\Big(\zaoa; \theta(\V{x}_n, \V{p}_{k,\text{va}}^{(j)}) \rmv\rmv + \rmv\rmv \eta_{k,n}^{(j)}, \sigma_{\theta}^{2} (\zu) \rmv\rmv \Big)   \s \mathrm{d} \eta_{\rmv  k,n}^{(j)} \nonumber \\	
	&= \frac{1}{2 {\psi}^{(j)}_{\theta,n}} \left( \text{erf}\left( \frac{\theta(\V{x}_n, \V{p}_{k,\text{va}}^{(j)}) + \psi^{(j)}_{\theta,n}/2 - \zaoa }{  \sigma_{\theta} (\zu) \sqrt{2}} \right) \right.
	\nonumber \\ &\hspace{13mm} - \left. \text{erf}\left( \frac{\theta(\V{x}_n, \V{p}_{k,\text{va}}^{(j)}) - \psi^{(j)}_{\theta,n}/2 - \zaoa }{  \sigma_{\theta} ( \zu) \sqrt{2}} \right) \right)
\label{eq:lhf_aoa_disp}\\[-7mm]\nn
\end{align}
where $f(\zaod | \V{p}_{n}, \V{p}_{k,\text{va}}^{(j)},\psi^{(j)}_{\vartheta,n})$ is defined in a similar manner as \eqref{eq:lhf_aoa_disp}. Note that in \eqref{eq:mainAoD}, we have to distinguish between $k=1$ and $k\neq 1$ (see Section~\ref{sec:signal_model}). It is also possible that a measurement $\RV{z}_{m,n}^{(j)}$ did not originate from any \ac{pva} (\emph{false positive} measurements comprise false alarms and clutter measurements). 
False positive measurements are assumed statistically independent of \ac{pva} states. 
They are modeled by a Poisson point process with mean $ {{\mu}_{\mathrm{fp}}} $ and \ac{pdf} $ f_{\mathrm{fp}}(\V{z}_{m,n}^{(j)}) $, which is assumed to factorize as $f_{\mathrm{fp}}(\V{z}_{m,n}^{(j)}) = {f_{\mathrm{fp}}}(\zd) {f_{\mathrm{fp}}}(\zaoa) {f_{\mathrm{fp}}}(\zaod)$. 
The false positive \ac{pdf} for a single delay measurement is assumed to be uniformly distributed as $ {f_{\mathrm{fp}}}(\zd)\rmv\rmv= f_\mathrm{U}(\zd;0,\tau_\text{max})$.
For a single \ac{aoa} and \ac{aod} measurement, the  false alarm \ac{pdf} is assumed to be distributed as  $ {f_{\mathrm{fp}}}(\zaoa)\rmv\rmv= f_\mathrm{U}(\zaoa;-\pi,\pi)$ and  $ {f_{\mathrm{fp}}}(\zaod)\rmv\rmv= f_\mathrm{U}(\zaod;-\pi,\pi)$.

The mean number of \ac{pva}-related measurements $\mu_\text{m}\big( \RV{x}^{(j)}_{k,n}\big) \triangleq \mu_\text{m}\big({\RV{\psi}}^{(j)}_{k,n}\big) $ is well approximated as  
\vspace{-2mm}
\begin{align} \label{eq:mean_number_of_meas}
	\mu_\text{m}\big({\RV{\psi}}^{(j)}_{k,n}\big)  = \Big(1 +\frac{N_{\text{ny}, \tau}\, \rv{\psi}^{(j)}_{\tau,n} }{c \, {T}_\text{s}}  + \frac{\rv{\psi}^{(j)}_{\theta,n}}{N_{\text{ny}, \theta}} + \frac{\rv{\psi}^{(j)}_{\vartheta,n}}{N_{\text{ny}, \vartheta}}\Big) p_{\text{d}} \\[-7mm]\nn
\end{align}
where $p_{\text{d}}$ is the probability of the detection associated with the detection threshold $\gamma$  of the \ac{ceda}, $N_{\text{ny}, \tau}$ is average number of components to be detected within one Nyquist sample, and $N_{\text{ny}, \theta}$ and $N_{\text{ny}, \vartheta}$ are the average numbers of components within the respective Rayleigh resolutions. 

\subsection{New \acp{pva}}

Newly detected \acp{pva}, i.e., actual \acp{va} that generate a measurement for the first time, are modeled by a Poisson point process with mean $\mu_\text{n}$ and PDF $f_\text{n}\big(\overline{\V{x}}^{(j)}_{m,n}|\V{x}_n\big)$. Following \cite{LeitMeyHlaWitTufWin:TWC2019,MeyerProc2018}, newly detected \acp{va} are represented by new \ac{pva} states $\overline{\RV{y}}^{(j)}_{m,n}$, $m \in \{1,\dots,\rv{M}_n^{(j)}\}$, where each new \ac{pva} state corresponds to a measurement $\RV{z}_{m,n}^{(j)}$; $\overline{r}_{m,n}^{(j)} \rreq 1$ implies that measurement  $\RV{z}_{m,n}^{(j)}$ was generated by a newly detected \ac{va}. 
Since newly detected \acp{va} can potentially produce more than one measurement, we use the multiple-measurement-to-feature probabilistic data association and define this mapping as introduced in \cite{MeyWilJ21,MeyerICASSP2020}. 
We also introduce the joint states $\overline{\RV{y}}_n^{(j)} \!\triangleq \big[\overline{\RV{y}}_{1,n}^{(j)\text{T}} \rmv\cdots\ist \overline{\RV{y}}_{M_{n}^{(j)}\rmv,n}^{(j)\text{T}} \big]^\T\rmv$ and $\overline{\RV{y}}_n \!\triangleq \big[\overline{\RV{y}}_n^{(1)\text{T}} \rmv\cdots\ist \overline{\RV{y}}_n^{(J)\text{T}} \big]^\T\rmv$. 
The vector of all \acp{pva} at time $n$ is given by $\RV{y}_{n} \!\triangleq \big[\underline{\RV{y}}_n^{\text{T}} \,\ist \overline{\RV{y}}_n^{\text{T}}\big]^\T\rmv$. 
Note that the total number of \acp{pva} per PA is given by $\rv{K}_{n}^{(j)} \rmv= K_{n-1}^{(j)} + \rv{M}_{n}^{(j)}$. 

Since new \acp{pva} are introduced as new measurements are available at each time, the number of \acp{pva} grows indefinitely. 
Thus, for feasible methods a suboptimal pruning step is employed that removes unlikely \acp{pva} (see Section~\ref{sec:problem}).

\subsection{Association Vectors}
\label{sec:assoc_vec_description}

\ac{slam} is complicated by data association uncertainty meaning that it is unknown which measurement $\V{z}^{(j)}_{m,n}$ originated from which \ac{va}. Furthermore, it is not known if a measurement did not originate from a PVA, i.e., it is a false positive measurement. Following \cite{MeyWilJ21}, we use measurement-oriented association variables
\vspace*{-1mm}
\begin{align}
	\rv{b}_{m,n}^{(j)} \triangleq\rmv \begin{cases} 
		k \in \{1,\dots,K_{n-1}^{(j)} + m \}, &\!\begin{minipage}[t]{45mm}if measurement $m$ was \\generated by \ac{pva} $k$\end{minipage}\\[5mm]
		0 \ist, &\!\begin{minipage}[t]{45mm} otherwise \end{minipage}
	\end{cases}\nn\\[-7mm]\nn
\end{align} 
and define the measurement-oriented association vector $\RV{b}^{(j)}_n = [\rv{b}_{1,n}^{(j)}\iist\cdots\iist\rv{b}_{\rv{M}_n^{(j)},n}^{(j)}]$. 
We also define $\RV{b}_{n} \triangleq [\RV{b}_{n}^{(1)\T}\iist\cdots\iist \RV{b}_{n}^{(J)\T}  ]^\T$. 

\subsection{Joint Posterior \ac{pdf}}
\label{sec:derivationFactorGraph}

\begin{figure}[!t]
\ifthenelse{\equal{\arxiv}{false}}
		{  
	\tikzsetnextfilename{FG_EOT_SLAM}
	\setlength{\abovecaptionskip}{-0.5mm}
	\setlength{\belowcaptionskip}{0pt}
	\hspace*{-3.2cm}\scalebox{0.95}{\input{plots/FG_EOT_SLAMv3.tex}} }
	{
	\includegraphics[scale=1]{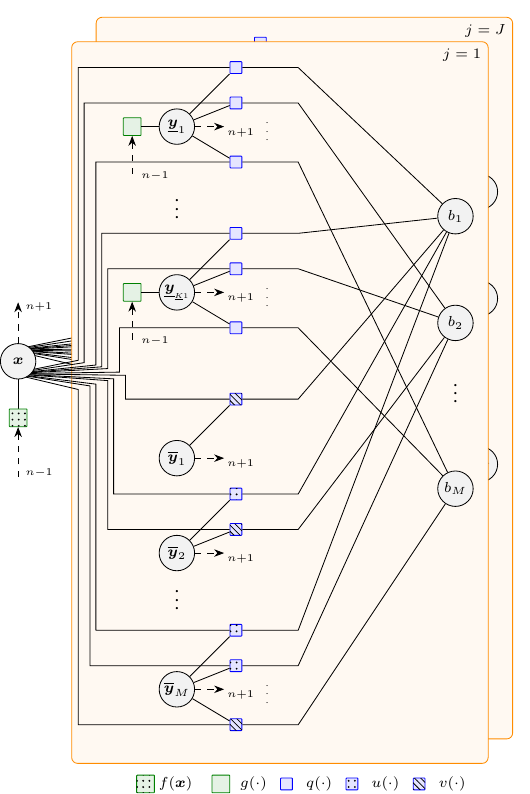} }
	\caption{Factor graph for proposed algorithm. At MP iteration $p$, we use the following short hand notation: $f(\V{x}) \triangleq f(\V{x}_n|\V{x}_{n-1})$, $g(\cdot)$, $q(\cdot)$, $u(\cdot)$ and $v(\cdot)$ corresponds to \eqref{eq:g}, \eqref{eq:q}, \eqref{eq:u}, \eqref{eq:v}, respectively. The time evolution of the agent state and \acp{va} is indicated with dashed arrows.}
	\label{fig:FG}
\end{figure}

By using common assumptions \cite{MeyerProc2018,LeitMeyHlaWitTufWin:TWC2019}, and for fixed (observed) measurements $\V{z}_{1:n}$, it can be shown that the joint posterior PDF of $\RV{x}_{1:n}$ ($\RV{x}_{1:n} \triangleq [\RV{x}_1^\T \cdots \RV{x}_n^\T]^\T$), $\RV{y}_{1:n}$, and $\RV{b}_{1:n}$, conditioned on $\V{z}_{1:n}$ for all time steps $n' \in \{1,\ist\dots\ist, n \}$ is given by
\vspace*{-2mm}
\begin{align}
\vspace*{-2mm}
&\hspace{-2mm}f( \V{x}_{1:n}, \V{y}_{1:n}, \V{b}_{1:n}| \V{z}_{1:n} ) \nn\\[-1.1mm]
&\hspace{-2mm}\propto  f(\V{x}_{1}) \prod^{J}_{j'=1} \rmv 
  \prod^{K^{(j')}_{1}}_{k'=1} \! f\big( \underline{\V{y}}^{(j')}_{k'\!,1}\big)   \prod^{n}_{n'=2}  \! f(\V{x}_{n'}|\V{x}_{n'-1}) \nn\\[-1.2mm]
&\hspace{0mm}\times\rmv\ \prod^{J}_{j=1}  \Bigg(\prod^{K^{(j)}_{n'-1}}_{k=1} g\big(\underline{\V{y}}^{(j)}_{k,n'} \big| \V{y}^{(j)}_{k,n'-1},\V{x}_{n'-1}\big)\nn\\[-1.3mm]
&\hspace{0mm}\times \prod^{M^{(j)}_{n'}}_{l=1} \rmv q\big( \V{x}_{n'}, \underline{\V{y}}^{(j)}_{k,n'}, b^{(j)}_{l,n'}; \V{z}^{(j)}_{l,n'} \big) \Bigg) \nn\\[-1mm]
&\hspace{0mm}\times \Bigg( \prod^{M^{(j)}_{n'}}_{m=1}  v\big( \V{x}_{n'}, \overline{\V{y}}^{(j)}_{m,n'}, b^{(j)}_{m,n'}; \V{z}^{(j)}_{m,n'} \big) \nn \\[-1mm]
& \hspace{0mm} \times\hspace{-1mm} \prod^{m-1}_{h=1} u\big( \V{x}_{n'}, \overline{\V{y}}^{(j)}_{m,n'}, b^{(j)}_{h,n'}; \V{z}^{(j)}_{h,n'} \big) \Bigg) 
\label{eq:factorization_post}\\[-7mm]\nn
\end{align}
where $g(\underline{\V{y}}^{(j)}_{k,n}|\underline{\V{y}}^{(j)}_{k,n-1},\xn{n-1})$, $q\big( \V{x}_{n}, \underline{\V{y}}^{(j)}_{k,n}, b^{(j)}_{l,n}; \V{z}^{(j)}_{l,n} \big)$, $u\big( \V{x}_{n}, \overline{\V{y}}^{(j)}_{k,n}, b^{(j)}_{h,n}; \V{z}^{(j)}_{h,n} \big)$ and $ v\big(  \V{x}_{n}, \overline{\V{y}}^{(j)}_{m,n}, b^{(j)}_{m,n}; \V{z}^{(j)}_{m,n} \big)$ are explained in what follows. 
The \emph{pseudo state-transition function} is given by
\begin{align}
  &\hspace*{-4mm}g(\underline{\V{y}}^{(j)}_{k,n}|\underline{\V{y}}^{(j)}_{k,n-1},\xn{n-1}) \nn\\
  &\hspace*{-2mm}\triangleq\rmv\rmv 
    \begin{cases}
      e^{-\mumL{k}{n-1}}  f(\underline{\V{x}}^{(j)}_{k,n},1 |\underline{\V{x}}^{(j)}_{k,n-1},\underline{r}^{(j)}_{k,n-1}), &\hspace*{-2mm} \underline{r}^{(j)}_{k,n} \rreq 1 \\[1mm]
      f(\underline{\V{x}}^{(j)}_{k,n},0 |\underline{\V{x}}^{(j)}_{k,n-1},\underline{r}^{(j)}_{k,n-1}), &\hspace*{-2mm} \underline{r}^{(j)}_{k,n} \rreq 0
    \end{cases} \label{eq:g}      
\end{align}
and the \emph{pseudo prior distribution} as
\vspace*{-1mm}
\begin{align}
\hspace*{-2.5mm}f(\overline{\V{y}}^{(j)}_{k,n}|\xn{n}) \rmv\rmv\triangleq \rmv\rmv
    \begin{cases}
      \mu_n f_n\big(\overline{\V{x}}^{(j)}_{k,n}|\V{x}_n\big) e^{-\mumN{k}{n}}, &\hspace*{-2mm}\overline{r}^{(j)}_{k,n} \rreq 1 \\[1mm]
      f_d\big(\overline{\V{x}}^{(j)}_{k,n}\big), &\hspace*{-2mm}\overline{r}^{(j)}_{k,n} \rreq 0 \ist.
    \end{cases} \label{eq:fy}     
\end{align}
The \emph{pseudo likelihood functions} related to legacy \acp{pva} for $k \in \Set{K}_{n-1}^{(j)}$ $q\big( \V{x}_n, \underline{\V{y}}^{(j)}_{k,n}, b^{(j)}_{l,n}; \V{z}^{(j)}_{l,n} \big) = q\big( \V{x}_n, \underline{\V{x}}^{(j)}_{k,n}, \underline{r}^{(j)}_{k}, b^{(j)}_{l,n}; \V{z}^{(j)}_{l,n} \big)$ is given by
\vspace*{-1mm}
\begin{align}
&\hspace*{-12mm} q\big( \V{x}_n, \underline{\V{x}}^{(j)}_{k,n}, 1, b^{(j)}_{l,n}; \V{z}^{(j)}_{l,n} \big) \nn\\
&\hspace*{5mm}\triangleq 
    \begin{cases}
      \frac{\mumL{k}{n} f (\V{z}^{(j)}_{l,n}| \V{p}_{n}, \underline{\V{x}}_{k,n}^{(j)})}{\mu_\text{fp} f_\text{fp}(\V{z}^{(j)}_{l,n})}, &  b^{(j)}_{l,n} \rreq  k\\
      1, &  b^{(j)}_{l,n} \rrneq k\\
    \end{cases}  \label{eq:q} 
\\[-7mm]\nn     
\end{align}
and $q\big( \V{x}_n, \underline{\V{x}}^{(j)}_{k,n}, 0, b^{(j)}_{l,n}; \V{z}^{(j)}_{l,n} \big) \triangleq 1-\delta(b^{(j)}_{l,n} - k)$. 
The \emph{pseudo likelihood functions} related to a new \ac{pva} (with $k \in \Set{M}_{n}^{(j)} \backslash m$) $u\big( \V{x}_n, \overline{\V{y}}^{(j)}_{k,n}, b^{(j)}_{h,n}; \V{z}^{(j)}_{h,n} \big) = u\big( \V{x}_n, \overline{\V{x}}^{(j)}_{k,n}, \overline{r}^{(j)}_{k}, b^{(j)}_{h,n}; \V{z}^{(j)}_{h,n} \big)$, where $h \in \{ 1,...,m-1\}$ is given by
\vspace*{-1mm}
\begin{align}
&\hspace*{-2mm} u\big( \V{x}_n, \overline{\V{x}}^{(j)}_{k,n}, 1, b^{(j)}_{h,n}; \V{z}^{(j)}_{h,n} \big) \nn\\
&\hspace*{-2mm}\triangleq 
    \begin{cases}
       \frac{f(\overline{\V{y}}^{(j)}_{k,n}|\xn{n}) \mumN{k}{n} f(\V{z}^{(j)}_{h,n}| \V{p}_{n}, \overline{\V{x}}_{k,n}^{(j)})}{\mu_\text{fp} f_\text{fp}(\V{z}^{(j)}_{h,n})}, & \rrmv \rrmv \rrmv b^{(j)}_{h,n} \rreq  K_{n-1}^{(j)} \rrmv+\rrmv k\\
     1, &  \rrmv \rrmv \rrmv b^{(j)}_{h,n} \rrneq K_{n-1}^{(j)} \rrmv+\rrmv k\\
    \end{cases}  \label{eq:u}  
\\[-7mm]\nn     
\end{align}
and $u\big( \V{x}_n, \overline{\V{x}}^{(j)}_{k,n}, 0, b^{(j)}_{h,n}; \V{z}^{(j)}_{h,n} \big) \rrmv  \triangleq \rrmv  1 \rrmv -\rrmv \delta(b^{(j)}_{h,n} \rrmv - \rrmv  (K_{n-1}^{(j)} \rrmv+\rrmv k))$, whereas for $k=m$ as $v\big(\V{x}_n, \overline{\V{y}}^{(j)}_{m}, b^{(j)}_{m,n}; \V{z}^{(j)}_{m,n} \big)  = v\big(\V{x}_n, \overline{\V{x}}^{(j)}_{m,n},\overline{r}^{(j)}_{m,n}, b^{(j)}_{m,n}; \V{z}^{(j)}_{m,n} \big)$ is given by
\vspace*{-1.5mm}
\begin{align}
  &v\big(\V{x}_n, \overline{\V{x}}^{(j)}_{m,n},1 , b^{(j)}_{m,n}; \V{z}^{(j)}_{m,n} \big) \nn\\
  &\hspace*{1mm}\triangleq 
    \begin{cases}
      \frac{f(\overline{\V{y}}^{(j)}_{m,n}|\xn{n}) \mumN{m}{n} f\big(\V{z}^{(j)}_{m,n}|\V{p}_{n},\overline{\V{x}}_{m,n}^{(j)}\big)}{\mu_\text{fp} f_\text{fp}(\V{z}^{(j)}_{m,n})}, & \rrmv \rrmv b^{(j)}_{m,n} \rreq K_{n-1}^{(j)} \rrmv+\rrmv m\\
      0,  & \rrmv \rrmv b^{(j)}_{m,n} \rrneq K_{n-1}^{(j)} \rrmv+\rrmv m\\
    \end{cases} \label{eq:v}  \\[-7.5mm]\nn
\end{align}
and $v\big(\V{x}_n, \overline{\V{x}}^{(j)}_{m,n},0 , b^{(j)}_{m,n}; \V{z}^{(j)}_{m,n} \big) \triangleq 1 \rrmv -\rrmv \delta(b^{(j)}_{m,n} \rrmv - \rrmv  (K_{n-1}^{(j)} \rrmv+\rrmv m))$. The factor graph representing the factorization \eqref{eq:factorization_post} is shown in Fig.~\ref{fig:FG}. The proposed message passing is similar to \cite{WieVenWilLei:JAIF2023_arxiv}.

\subsection{Detection of \acp{pva} and State Estimation}\label{sec:problem}

At each time $n$ and for each \ac{pa} $j$, the \ac{ceda} provides the currently observed measurement vector $\V{z}_n^{(j)}$, with fixed ${M}^{(j)}_n$, according to Section~\ref{sec:channel_estimation}. We aim to estimate all states using all available measurements $\V{z}_{1:n}\rmv$ from all \acp{pa} up to time $n$. We calculate estimates of the agent state $\RV{x}_n$ by using the \ac{mmse} estimator \vspace{.5mm} \cite[Ch.~4]{Poo:B94}, i.e., 
\vspace*{-2mm}
\begin{align}
	\V{x}^{\text{MMSE}}_{n} &\triangleq\, \int \rmv \V{x}_n \ist f( \V{x}_n|\V{z}_{1:n}) \ist \mathrm{d} \V{x}_n \label{eq:MMSEagent}\\[-7mm]\nn
\end{align}
The map of the environment is represented by reflective surfaces described by \acp{pva}. 
Therefore, the state $\V{x}_{k,n}^{(j)}$ of the detected \acp{pva} $k \!\in\! \{ 1,\dots,K^{(j)}_n \}$ must be estimated. 
This relies on the marginal posterior existence probabilities $p(r^{(j)}_{k,n} \!=\! 1|\V{z}_{1:n}) = \int f(\V{x}_{k,n}^{(j)} , r^{(j)}_{k,n} \!=\! 1| \V{z}^{(j)}_{1:n} ) \mathrm{d}\V{x}_{k,n}^{(j)}$ and the marginal posterior \acp{pdf} $f(\V{x}_{k,n}^{(j)}| r^{(j)}_{k,n} \!=\! 1, \V{z}_{1:n} ) \rmv\rmv=\rmv\rmv f(\V{x}_{k,n}^{(j)}, r^{(j)}_{k,n} \!=\! 1| \V{z}_{1:n} )/p(r^{(j)}_{k,n} \!=\! 1|\V{z}_{1:n})$. 
A \ac{pva} $k$ is declared to exist if $p(r^{(j)}_{k,n} \!=\! 1|\V{z}_{1:n}) > p_{\text{cf}}$, where $p_{\text{cf}}$ is a confirmation threshold \cite[Ch.~2]{Poo:B94}. 
To avoid that the number of \ac{pva} states grows indefinitely, \ac{pva} states with $p(r^{(j)}_{k,n} \!=\! 1|\V{z}_{1:n})$ below a threshold $p_{\text{pr}}$ are removed from the state space (``pruned''). 
The number $\hat{K}^{(j)}_n$ of \ac{pva} states that are considered to exist is the estimate of the total number $L_n^{(j)}$ of \acp{va} visible at time $n$. 
For existing \acp{pva}, an estimate of its state $\RV{x}_{k,n}^{(j)}$ can again be calculated by the \ac{mmse}
\vspace*{-1.5mm}
\begin{align}
	\V{x}_{k,n}^{(j)\ist\text{MMSE}}  \,\triangleq \int \rmv \V{x}_{k,n}^{(j)}  \ist\ist f(\V{x}_{k,n}^{(j)}\ist | \ist r^{(j)}_{k,n} \!=\! 1, \V{z}_{1:n}) \ist\ist \mathrm{d}\V{x}_{k,n}^{(j)} \rmv \label{eq:MMSEpva}\\[-7mm]\nn
\end{align}
where $\V{x}_{k,n}^{(j)\ist\text{MMSE}} = [\V{p}_{k,n}^{(j)\ist\text{MMSE}},\V{\psi}_{k,n}^{(j)\ist\text{MMSE}}]^\T$. The calculation of $f( \V{x}_n|\V{z}_{1:n})$, $p(r_{k,n} \rmv\rmv=\rmv\rmv 1 |\V{z})$, and $f(\V{x}_{k,n}^{(j)} |$ $r^{(j)}_{k,n} \rmv\rmv=\rmv\rmv 1, \V{z}_{1:n})$ from the joint posterior $f(\V{x}_{1:n}, \V{y}_{1:n},\V{b}_{1:n}| \V{z}_{1:n} )$ by direct marginalization is not feasible. 
By performing sequential particle-based \ac{MP} \cite{MeyHliHla:TSPIN2016, MeyerProc2018,LeitMeyHlaWitTufWin:TWC2019} using the SPA rules \cite{KscFreLoe:TIT2001} on the factor graph in Fig.~\ref{fig:FG}, approximations (``beliefs'') $b\big(\V{x}_{n} \big)$ and $b\big(\V{y}^{(j)}_{k,n} \big)$ of the marginal posterior \acp{pdf} $f( \V{x}_n|\V{z}_{1:n})$, $p(r^{(j)}_{k,n} \!=\! 1 |\V{z}_{1:n})$, and $f(\V{x}_{k,n}^{(j)}|$ $r^{(j)}_{k,n} \rmv=\rmv 1, \V{z}_{1:n})$ can be obtained in an efficient way for the agent state as well as all legacy and new \ac{pva} states.


\section{Numerical Results}
\label{sec:results}

\begin{figure}[!t]
	\ifthenelse{\equal{\arxiv}{false}}
	{  
		\tikzsetnextfilename{floorplan}
		\centering
		\setlength{\abovecaptionskip}{-0.5mm}
		\setlength{\belowcaptionskip}{0pt}
		\scalebox{1}{\input{plots/floorplan_rotated.tikz}}}
	{
	\includegraphics[scale=1]{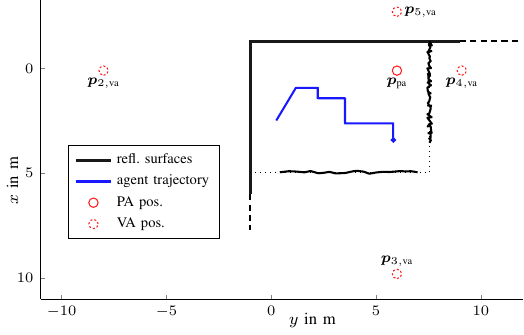}}
	\caption{Considered scenario for performance evaluation. Perfect reflectors are indicated with straight lines and rough surfaces with wavy lines.}
	\label{fig:floorplan}
	\vspace*{-5mm}
\end{figure}

We consider an indoor scenario shown in Fig.~\ref{fig:floorplan}. 
The scenario consists of one \ac{pa} at position $\V{p}_{\mathrm{pa}}^{(1)} = [0.1 \iist 6]^{\T}$ and four reflective surfaces, i.e., $4$ \acp{va}. The agent moves along a track which is observed for $300$ time instances $n$ with observation period $\Delta T = 1\,$s. 

\def\figHresults{0.23\columnwidth}
\def\figWresults{0.55\columnwidth}
\begin{figure*}[!t]
	\setlength{\abovecaptionskip}{-0.5mm}
	\setlength{\belowcaptionskip}{0pt}
	\captionsetup[subfigure]{captionskip=0pt}
	\centering
	\ifthenelse{\equal{\arxiv}{false}}
	{ 
		\subfloat[]{\tikzsetnextfilename{RMSEoverTime}\hspace{8mm}
			\input{plots/RMSEoverTime.tex}\label{fig:rmseVsTime}}
		\subfloat[]{\tikzsetnextfilename{MOSPA_PA1}\hspace{8mm}
			\input{plots/MOSPA_PA1.tex}\label{fig:mospaPA1}}
		\subfloat[]{\tikzsetnextfilename{CardError_PA1}\hspace{8mm}
			\input{plots/CardError_PA1.tex}\label{fig:cardErrorPA1}} \\\vspace{-2mm}
		\subfloat[]{\tikzsetnextfilename{Extent_ed}\hspace{9mm}
			\input{plots/Extent_ed.tex}\label{fig:extend_d}}
		\subfloat[]{\tikzsetnextfilename{Extent_aoa}\hspace{9.7mm}
			\input{plots/Extent_aoa.tex}\label{fig:extend_aoa}} 
		\subfloat[]{\tikzsetnextfilename{Extent_aod}\hspace{8mm}
			\input{plots/Extent_aod.tex}\label{fig:extend_aod}}\\[1mm]
		\hspace*{0cm}
		\subfloat{\tikzsetnextfilename{legend_Ex1} \legendref{nameS}\hfill}\\
	}
	{
		\subfloat[]{\hspace{3mm}
			\includegraphics{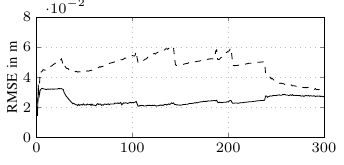}\label{fig:rmseVsTime}}
		\subfloat[]{\hspace{-3mm}
			\includegraphics{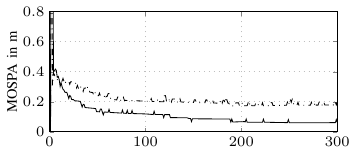}\label{fig:mospaPA1}}
		\subfloat[]{\hspace{-3mm}
			\includegraphics{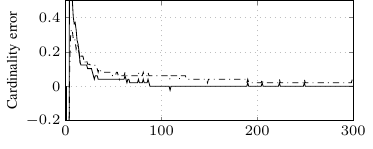}\label{fig:cardErrorPA1}} \\\vspace{-6mm}
		\subfloat[]{\hspace{3mm}
			\includegraphics{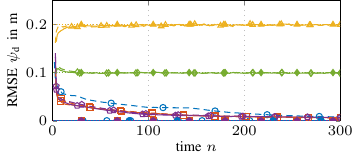}\label{fig:extend_d}}
		\subfloat[]{\hspace{1.5mm}
			\includegraphics{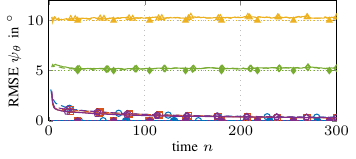}\label{fig:extend_aoa}} 
		\subfloat[]{\hspace{-1.5mm}
			\includegraphics{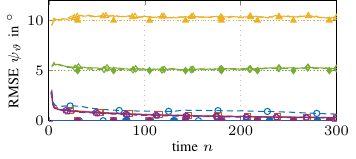}\label{fig:extend_aod}}\\[1mm]
		\hspace*{0cm}
		\subfloat{\includegraphics{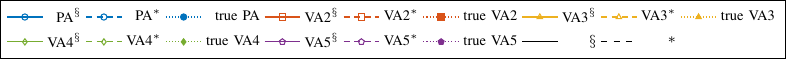}\hfill}\\
	}
	\vspace*{1mm}
	\caption{Results for converged simulation runs. (a) shows the \ac{rmse} of the agent position over the whole track. (b) and (c) present the \ac{mospa} and the cardinality error. (d) - (f) show the  \ac{rmse} of the dispersion parameters. The legend is explained in what follows. The word \textit{true} and dotted lines indicate the true dispersion parameters for each PA or VA. $§$ and solid lines correspond to unknown VA dispersion parameters but known PA dispersion parameters.  $*$ and dashed lines indicate unknown dispersion parameters.}
	\label{fig:results}
	\vspace*{-4mm}
\end{figure*}

Measurements are generated according to the proposed system model in Section~\ref{sec:measurementModel} and to the scenario shown in Fig.~\ref{fig:floorplan}. The distances of the main components are calculated based on the \ac{pa} and the corresponding \acp{va} positions as well as the agent positions (see Section~\ref{sec:signal_model}). The signal \ac{snr} is set to {$40\,\mathrm{dB}$} at an \ac{los} distance of $1\,$m. The amplitudes of the main-components (\ac{los} component and the \acp{mpc}) are calculated using a free-space path loss model and an additional attenuation of {$3\,\mathrm{dB}$} for each reflection at a surface.
The sub-components are generated with a constant amplitude attenuation $\beta_{l,i,n} = 0.9$ for all $i$ and for all $n$ and fixed dispersion parameters ${\psi}_{\tau,l,n} = \psi_{\tau,l} = \psi_{\text{d},l}/c$, ${\psi}_{\theta,k,n} = {\psi}_{\theta,k}$, and ${\psi}_{\vartheta,k,n} = {\psi}_{\vartheta,k}$ for all $n$. We define $ N_\text{ny} = 4$, $ N_{\text{ny}, \theta} =  N_{\text{ny}, \vartheta} = 2$ and $p_\text{d} = 0.98$. In addition false positive measurements are generated according to the model in Section~\ref{sec:measurementModel} with a mean number of $\mu_\text{fp} = 5$. For the calculation of the measurement variances, we assume a $3\text{-}\mathrm{dB}$ system bandwidth of $B=500\,\mathrm{MHz}$ for some known transmit signal spectrum for a carrier frequency of $f_\mathrm{c} = 6\,\mathrm{GHz}$. The arrays employed at agent and \ac{pa} are of identical geometry with $H = H^{(j)}=9$ antenna elements in (known) \ac{ura}-configuration spaced at $\lambda/2$, where $\lambda=c/f_\mathrm{c}$ is the carrier wavelength.
We use $10^5$ particles. The particles for the initial agent state are drawn from a 4-D uniform distribution with center $\V{x}_0 = [\V{p}_{0}^{\T}\;0\;\, 0]^{\T}\rmv$, where $\V{p}_{0}$ is the starting position of the actual agent track, and the support of each position component about the respective center is given by $[-0.1\,\mathrm{m}, 0.1\,\mathrm{m}]$ and of each velocity component is given by $[-0.01\,\mathrm{m/s}, 0.01\,\mathrm{m/s}]$. 
At time $n \rmv=\rmv 0$, the number of \acp{va} is $0$, i.e., no prior map information is available. 
The prior distribution for new \ac{pva} states $f_\text{n}\big(\overline{\V{x}}^{(j)}_{m,n}|\V{x}_n\big)$ is uniform on the square region given by $[-\text{15 m}, \text{15 m}] \times [-\text{15 m}, \text{15 m}]$ around the center of the floor plan shown in Fig.~\ref{fig:floorplan} and the mean number of new \acp{pva} at time $n$ is  $\mu_\text{n} = 0.01$. 
The probability of survival is $p_{\mathrm{s}} = 0.999$. The confirmation threshold as well as the pruning threshold are given as $p_{\mathrm{cf}} = 0.5$ and $p_{\mathrm{pr}} = 10^{-3}$, respectively. For the sake of numerical stability, we introduce a small amount of regularization noise to the \ac{va} state $\V{p}_{k,\mathrm{va}}$ at each time step $n$, i.e., $\underline{\V{p}}^{(j)}_{k,\mathrm{va}} \rmv\rmv=\rmv\rmv \V{p}^{(j)}_{k,\mathrm{va}} \rmv+\rmv \V{\omega}_{k}$, where $\V{\omega}_{k}$ is \ac{iid} across $k$, zero-mean, and Gaussian with covariance matrix $\sigma_a^2\, \V{I}_2$ and $\sigma_a = 10^{-3}\,\text{m}$. 
The state transition variances are set as $\sigma_\text{w} = 10^{-3}\,\mathrm{m/s^2}$, $q_{\tau} = q_{\theta} = q_{\vartheta} = 10^{3}$ \cite{Koc:TAES2008_EOT,MeyWilJ21}. 
The performance is measured in terms of the \ac{rmse} of the agent position and the dispersion parameters as well as the \acf{ospa} error \cite{Schuhmacher2008} of all \acp{va} with cutoff parameter and order set to 5~m and 2, respectively. 
The \ac{mospa} errors and \acp{rmse} of each unknown variable are obtained by averaging over all converged simulation runs. 

\subsubsection*{Experiment}
\label{sec:ResultsModel}

For the investigation of the proposed method, we define for each surface an individual roughness as shown in Fig.~\ref{fig:floorplan}. This is captured by individual dispersion parameter settings as given in Table~\ref{tb:psi_settings}. We performed $100$ simulation runs. In each simulation run, we generated noisy measurements $\V{z}_{m,n}^{(j)}$ according to the measurement model proposed in Section~\ref{sec:measurementModel}. In the case $\psi_{k} = 0$ only main-component measurements are generated, which is equivalent to the system model in \cite{LeiGreWit:ICC2019}.
\begin{table}[!t]
	\caption{Dispersion parameter settings}
	\begin{center}
		\begin{tabular}{ c| l| l| l| l| l}
			&PA & VA1 &VA2 &VA3 & VA4 \\
			\hline \hline
			 $\psi_\text{d} $ & 0~m & 0~m & 0.2~m & 0.1~m & 0~m\\ \hline
			$\psi_\theta $ & 0$^\circ$ & 0$^\circ$  & 10$^\circ$  & 5$^\circ$  & 0$^\circ$  \\ \hline
			$\psi_\vartheta $ & 0$^\circ$  & 0$^\circ$  & 10$^\circ$  & 5$^\circ$  & 0$^\circ$  \\
 \hline \hline
		\end{tabular}
	\end{center}
	\label{tb:psi_settings}
	\vspace*{-7mm}
\end{table} 
The results are summarized in Fig.~\ref{fig:results}. We investigated the impact of two different settings on the estimation performance. The first one assumes the dispersion parameters of the PA as known and of the VAs as unknown. The second one assumes all dispersion parameters as unknown.
The \ac{rmse} of the agent positions is shown in Fig.~\ref{fig:rmseVsTime}. Assuming known dispersion parameters for the PA results in the smallest RMSE. The results show that estimating the dispersion parameters comes at the cost of localization accuracy.
Fig.~\ref{fig:mospaPA1} and \ref{fig:cardErrorPA1} show the \ac{mospa} error and its mean cardinality error contributions, respectively. After a few time steps, the number of VAs is estimated correctly, hence the only significant contribution in the \ac{mospa} are the positioning errors of the VAs. It shows a slight performance gap between unknown and known PA dispersion parameters due to the same reasons as mentioned above.
The \ac{rmse}s of the dispersion parameters are presented in Fig.~\ref{fig:extend_d}, \ref{fig:extend_aoa} and \ref{fig:extend_aod}. The true values are indicated with dashed lines in the same color. The results show that the dispersion parameters are well estimated and converge to the true value. For dispersion parameters equal to zero, the convergence is much slower. This is due to the fact that for a single measurement, there is a model mismatch in the likelihood. This could lead to a biased estimate at the beginning, which reduces to zero over time as indicated by the results. In addition, knowing the PA dispersion parameters has no significant impact on the estimation accuracy of the other dispersion parameters. For the unknown PA dispersion parameters, we observe a slightly slower convergence to zero. We assume that this is caused by the smaller measurement variance due to the higher amplitude of the \ac{los} path.

\section{Conclusions}\label{sec:conclusion}
We have proposed a new \ac{Mpslam} for non-ideal reflective surfaces in MIMO wireless communication systems where single- or multiple-measurements can be generated by a single environment feature (i.e., a single \ac{va}). By introducing individual dispersion parameters for each PA and VA, the proposed method fuses the information contained in ideal and non-ideal reflective surfaces. 
We demonstrated that the proposed algorithm can robustly and jointly estimate the position and dispersion extents of ideal and non-ideal reflective in a challenging scenario involving different types of surfaces.
Possible directions for future research include incorporating multiple-measurements-to-feature data association into the \ac{mva}-based \ac{slam} method.


\renewcommand{\baselinestretch}{0.96}\small\normalsize 
\bibliographystyle{IEEEtran}
\bibliography{IEEEabrv,references}

\begin{thebibliography}{10}
\providecommand{\url}[1]{#1}
\csname url@samestyle\endcsname
\providecommand{\newblock}{\relax}
\providecommand{\bibinfo}[2]{#2}
\providecommand{\BIBentrySTDinterwordspacing}{\spaceskip=0pt\relax}
\providecommand{\BIBentryALTinterwordstretchfactor}{4}
\providecommand{\BIBentryALTinterwordspacing}{\spaceskip=\fontdimen2\font plus
\BIBentryALTinterwordstretchfactor\fontdimen3\font minus
  \fontdimen4\font\relax}
\providecommand{\BIBforeignlanguage}[2]{{%
\expandafter\ifx\csname l@#1\endcsname\relax
\typeout{** WARNING: IEEEtran.bst: No hyphenation pattern has been}%
\typeout{** loaded for the language `#1'. Using the pattern for}%
\typeout{** the default language instead.}%
\else
\language=\csname l@#1\endcsname
\fi
#2}}
\providecommand{\BIBdecl}{\relax}
\BIBdecl

\bibitem{LeitingerJSAC2015}
E.~Leitinger, P.~Meissner, C.~Rudisser, G.~Dumphart, and K.~Witrisal,
  ``Evaluation of position-related information in multipath components for
  indoor positioning,'' \emph{{IEEE} J. Sel. Areas Commun.}, vol.~33, no.~11,
  pp. 2313--2328, Nov. 2015.

\bibitem{WitrisalSPM2016}
K.~Witrisal, P.~Meissner, E.~Leitinger, Y.~Shen, C.~Gustafson, F.~Tufvesson,
  K.~Haneda, D.~Dardari, A.~F. Molisch, A.~Conti, and M.~Z. Win,
  ``High-accuracy localization for assisted living: {5G} systems will turn
  multipath channels from foe to friend,'' \emph{{IEEE} Signal Process. Mag.},
  vol.~33, no.~2, pp. 59--70, Mar. 2016.

\bibitem{LeitMeyHlaWitTufWin:TWC2019}
E.~{Leitinger}, F.~{Meyer}, F.~{Hlawatsch}, K.~{Witrisal}, F.~{Tufvesson}, and
  M.~Z. {Win}, ``A belief propagation algorithm for multipath-based {SLAM},''
  \emph{{IEEE} Trans. Wireless Commun.}, vol.~18, no.~12, pp. 5613--5629, Dec.
  2019.

\bibitem{MenMeyBauWin:JSTSP2019}
R.~{Mendrzik}, F.~{Meyer}, G.~{Bauch}, and M.~Z. {Win}, ``Enabling situational
  awareness in millimeter wave massive {MIMO} systems,'' \emph{{IEEE} J. Sel.
  Topics Signal Process.}, vol.~13, no.~5, pp. 1196--1211, Sep. 2019.

\bibitem{GentnerTWC2016}
C.~Gentner, W.~Jost, T.and~Wang, S.~Zhang, A.~Dammann, and U.~C. Fiebig,
  ``Multipath assisted positioning with simultaneous localization and
  mapping,'' \emph{{IEEE} Trans. Wireless Commun.}, vol.~15, no.~9, pp.
  6104--6117, Sept. 2016.

\bibitem{DurrantWhyte2006}
H.~Durrant-Whyte and T.~Bailey, ``{Simultaneous localization and mapping: Part
  I},'' \emph{IEEE Robot. Autom. Mag.}, vol.~13, no.~2, pp. 99--110, June 2006.

\bibitem{Dissanayake2001}
M.~Dissanayake, P.~Newman, S.~Clark, H.~Durrant-Whyte, and M.~Csorba, ``A
  solution to the simultaneous localization and map building ({SLAM})
  problem,'' \emph{IEEE Trans. Robot. Autom.}, vol.~17, no.~3, pp. 229--241,
  June 2001.

\bibitem{LeiGreWit:ICC2019}
E.~{Leitinger}, S.~{Grebien}, and K.~{Witrisal}, ``Multipath-based {SLAM}
  exploiting {AoA} and amplitude information,'' in \emph{Proc. IEEE ICCW-19},
  Shanghai, China, May 2019, pp. 1--7.

\bibitem{KimGraGaoBatKimWym:TWC2020}
H.~{Kim}, K.~{Granstr{\"o}m}, L.~{Gao}, G.~{Battistelli}, S.~{Kim}, and
  H.~{Wymeersch}, ``{5G} {mmWave} cooperative positioning and mapping using
  multi-model {PHD} filter and map fusion,'' \emph{{IEEE} Trans. Wireless
  Commun.}, vol.~19, no.~6, pp. 3782--3795, Mar. 2020.

\bibitem{KimGranSveKimWym:TVT2022}
H.~Kim, K.~Granstrom, L.~Svensson, S.~Kim, and H.~Wymeersch, ``{PMBM-based
  SLAM} filters in {5G} {mmWave} vehicular networks,'' \emph{{IEEE} Trans. Veh.
  Technol.}, pp. 1--1, May 2022.

\bibitem{LeiMey:Asilomar2020_DataFusion}
E.~Leitinger and F.~Meyer, ``Data fusion for multipath-based {SLAM},'' in
  \emph{Proc. Asilomar-20}, Pacifc Grove, CA, USA, Oct. 2020, pp. 934--939.

\bibitem{LeiVenTeaMey:TSP2023}
E.~{Leitinger}, A.~{Venus}, B.~{Teague}, and F.~{Meyer}, ``Data fusion for
  multipath-based {SLAM}: {Combining} information from multiple propagation
  paths,'' \emph{{IEEE} Trans. Signal Process.}, vol.~71, pp. 4011--4028, Sep.
  2023.

\bibitem{ShutWanJos:CSTA2013}
D.~Shutin, W.~Wang, and T.~Jost, ``Incremental sparse {B}ayesian learning for
  parameter estimation of superimposed signals,'' in \emph{Proc. SAMPTA-2013},
  no.~1, Sept. 2013, pp. 6--9.

\bibitem{BadHanFle:TSP2017}
M.~A. Badiu, T.~L. Hansen, and B.~H. Fleury, ``Variational {Bayesian} inference
  of line spectra,'' \emph{{IEEE} Trans. Signal Process.}, vol.~65, no.~9, pp.
  2247--2261, May 2017.

\bibitem{LiLeiVenTuf:TWC2022}
X.~Li, E.~Leitinger, A.~Venus, and F.~Tufvesson, ``Sequential detection and
  estimation of multipath channel parameters using belief propagation,''
  \emph{{IEEE} Trans. Wireless Commun.}, vol.~21, no.~10, pp. 8385--8402, Apr.
  2022.

\bibitem{GreLeiWitFle:TWC2024}
S.~Grebien, E.~Leitinger, K.~Witrisal, and B.~H. Fleury, ``Super-resolution
  estimation of {UWB} channels including the dense component -- {An
  SBL}-inspired approach,'' \emph{{IEEE} Trans. Wireless Commun.}, 2024.

\bibitem{MeyWilJ21}
F.~Meyer and J.~L. Williams, ``Scalable detection and tracking of geometric
  extended objects,'' \emph{{IEEE} Trans. Signal Process.}, vol.~69, pp.
  6283--6298, Oct. 2021.

\bibitem{KulmerPIMRC2018}
J.~{Kulmer}, F.~{Wen}, N.~{Garcia}, H.~{Wymeersch}, and K.~{Witrisal}, ``Impact
  of rough surface scattering on stochastic multipath component models,'' in
  \emph{Proc. IEEE PIMRC 2018}, Bologna, Italy, Dec. 2018, pp. 1410--1416.

\bibitem{WenKulWitWym:TWC2021}
F.~Wen, J.~Kulmer, K.~Witrisal, and H.~Wymeersch, ``5{G} positioning and
  mapping with diffuse multipath,'' \emph{{IEEE} Trans. Wireless Commun.},
  vol.~20, no.~2, pp. 1164--1174, 2021.

\bibitem{WieVenWilLei:JAIF2023_arxiv}
L.~Wielandner, A.~Venus, T.~Wilding, and E.~Leitinger, ``Multipath-based {SLAM}
  for non-ideal reflective surfaces exploiting multiple-measurement data
  association,'' \emph{arXiv preprint arXiv:2304.05680}, 2023.

\bibitem{WieVenWilLei:Fusion2023}
------, ``Multipath-based {SLAM} with multiple-measurement data association,''
  in \emph{Proc. Fusion-23}, Charleston, USA, Jul. 2023, pp. 1--8.

\bibitem{MeyerICASSP2020}
F.~{Meyer} and J.~L. {Williams}, ``Scalable detection and tracking of extended
  objects,'' in \emph{Proc. ICASSP 2020}, Barcelona, Spain, May 2020, pp.
  8916--8920.

\bibitem{Kay1998}
S.~M. Kay, \emph{Fundamentals of Statistical Signal Processing: {D}etection
  Theory}.\hskip 1em plus 0.5em minus 0.4em\relax Upper Saddle River, NJ, USA:
  Prentice Hall, 1998.

\bibitem{SalVal:JSAC1987}
A.~Saleh and R.~Valenzuela, ``A statistical model for indoor multipath
  propagation,'' \emph{{IEEE} J. Sel. Areas Commun.}, vol.~5, no.~2, pp.
  128--137, Feb. 1987.

\bibitem{PedersenJTAP2018}
T.~Pedersen, ``Modeling of path arrival rate for in-room radio channels with
  directive antennas,'' \emph{{IEEE} Trans. Antennas Propag.}, vol.~66, no.~9,
  pp. 4791--4805, 2018.

\bibitem{MeyerProc2018}
F.~Meyer, T.~Kropfreiter, J.~L. Williams, R.~Lau, F.~Hlawatsch, P.~Braca, and
  M.~Z. Win, ``Message passing algorithms for scalable multitarget tracking,''
  \emph{Proc. {IEEE}}, vol. 106, no.~2, pp. 221--259, Feb. 2018.

\bibitem{BarShalom2002EstimationTracking}
Y.~Bar-Shalom, T.~Kirubarajan, and X.-R. Li, \emph{Estimation with Applications
  to Tracking and Navigation}.\hskip 1em plus 0.5em minus 0.4em\relax New York,
  NY, USA: John Wiley \& Sons, Inc., 2002.

\bibitem{Koc:TAES2008_EOT}
J.~W. {Koch}, ``Bayesian approach to extended object and cluster tracking using
  random matrices,'' \emph{{IEEE} Trans. Aerosp. Electron. Syst.}, vol.~44,
  no.~3, pp. 1042--1059, Jul. 2008.

\bibitem{WilGreLeiMueWit:ACSSC2018}
T.~Wilding, S.~Grebien, E.~Leitinger, U.~M\"uhlmann, and K.~Witrisal,
  ``Single-anchor, multipath-assisted indoor positioning with aliased antenna
  arrays,'' in \emph{Proc. Asilomar-18}, Pacifc Grove, CA, USA, Oct. 2018, pp.
  525--531.

\bibitem{Poo:B94}
H.~V. Poor, \emph{An Introduction to Signal Detection and Estimation},
  2nd~ed.\hskip 1em plus 0.5em minus 0.4em\relax New York: Springer-Verlag,
  1994.

\bibitem{MeyHliHla:TSPIN2016}
F.~Meyer, O.~Hlinka, H.~Wymeersch, E.~Riegler, and F.~Hlawatsch, ``Distributed
  localization and tracking of mobile networks including noncooperative
  objects,'' vol.~2, no.~1, pp. 57--71, Mar. 2016.

\bibitem{KscFreLoe:TIT2001}
F.~Kschischang, B.~Frey, and H.-A. Loeliger, ``Factor graphs and the
  sum-product algorithm,'' \emph{{IEEE} Trans. Inf. Theory}, vol.~47, no.~2,
  pp. 498--519, Feb. 2001.

\bibitem{Schuhmacher2008}
D.~Schuhmacher, B.-T. Vo, and B.-N. Vo, ``{A consistent metric for performance
  evaluation of multi-object filters},'' \emph{{IEEE} Trans. Signal Process.},
  vol.~56, no.~8, pp. 3447--3457, Aug. 2008.

\end{thebibliography}
\end{document}